\documentclass[pdflatex,sn-standardnature]{sn-jnl}

\jyear{2022}%

\theoremstyle{thmstyleone}%
\theoremstyle{thmstyletwo}%

\theoremstyle{thmstylethree}%

\usepackage{enumitem}
\usepackage{graphicx}
\usepackage{subcaption}
\usepackage{url}
\usepackage{amsmath}
\usepackage{amssymb}
\usepackage{setspace}
\usepackage{adjustbox}

\newcommand{\BI}{\begin{itemize}}
\newcommand{\EI}{\end{itemize}}
\newcommand{\oiii}{O\,{\sc iii}}
\newcommand{\oii}{O\,{\sc ii}}

\begin{document}

\title[\textcolor{white}{X}]{\Large Dilution of chemical enrichment in galaxies 600 Myr after the Big Bang}

\author[1,2]{Kasper~E. Heintz}
\author[1,2]{Gabriel~B. Brammer}
\author[1,2]{Clara Giménez-Arteaga}
\author[1,2]{Victoria~B. Strait}
\author[3,4,1]{Claudia del P. Lagos}
\author[1,5]{Aswin~P. Vijayan}
\author[6]{Jorryt Matthee}
\author[1,2]{Darach Watson}
\author[1,2]{Charlotte~A. Mason}
\author[1,2]{Anne Hutter}
\author[1,2]{Sune Toft}
\author[1,2]{Johan~P.~U. Fynbo}
\author[7,1,2]{Pascal~A. Oesch}

\affil[1]{\small Cosmic Dawn Center (DAWN), Denmark}
\affil[2]{\small Niels Bohr Institute, University of Copenhagen, Jagtvej 128, DK-2200 Copenhagen N, Denmark}
\affil[3]{\small International Centre for Radio Astronomy Research (ICRAR), M468, University of Western Australia, 35 Stirling Hwy, Crawley, WA
6009, Australia}
\affil[4]{\small ARC of Excellence for All Sky Astrophysics in 3 Dimensions (ASTRO 3D)}
\affil[5]{\small DTU-Space, Technical University of Denmark, Elektrovej 327, 2800, Kgs. Lyngby, Denmark}
\affil[6]{\small Department of Physics, ETH Zürich, Wolfgang-Pauli-Strasse 27, Zürich, 8093, Switzerland}
\affil[7]{\small Observatoire de Genève, Université de Genève, Chemin Pegasi 51, CH-1290 Versoix, Switzerland}


\abstract{\bf Galaxies throughout the last 12 Gyr of cosmic time follow a single, universal relation that connects their star-formation rates (SFRs), stellar masses ($M_\star$) and chemical abundances \cite{Mannucci10,Curti20}. Deviation from these fundamental scaling relations would imply a drastic change in the processes that regulate galaxy evolution. Observations have hinted at the possibility that this relation may be broken in the very early universe \cite{Troncoso14,Onodera16}. 
However, until recently, chemical abundances of galaxies could be only measured reliably as far back as redshift $z=3.3$ \cite{Sanders21}.
With JWST, we can now characterize the SFR, $M_\star$, and chemical abundance of galaxies during the first few hundred million years after the Big Bang, at redshifts $z= 7-10$. 
Here we show that galaxies at this epoch follow unique SFR-$M_\star$--main-sequence and mass-metallicity scaling relations, but their chemical abundance is a factor of three lower than expected from the fundamental-metallicity relation of later galaxies. 
These findings suggest that galaxies at this time are still intimately connected with the intergalactic medium and subject to continuous infall of pristine gas which effectively dilutes their metal abundances. 
}


\maketitle

\section*{Main text}

\begin{figure}
\centering
\includegraphics[width=\textwidth]{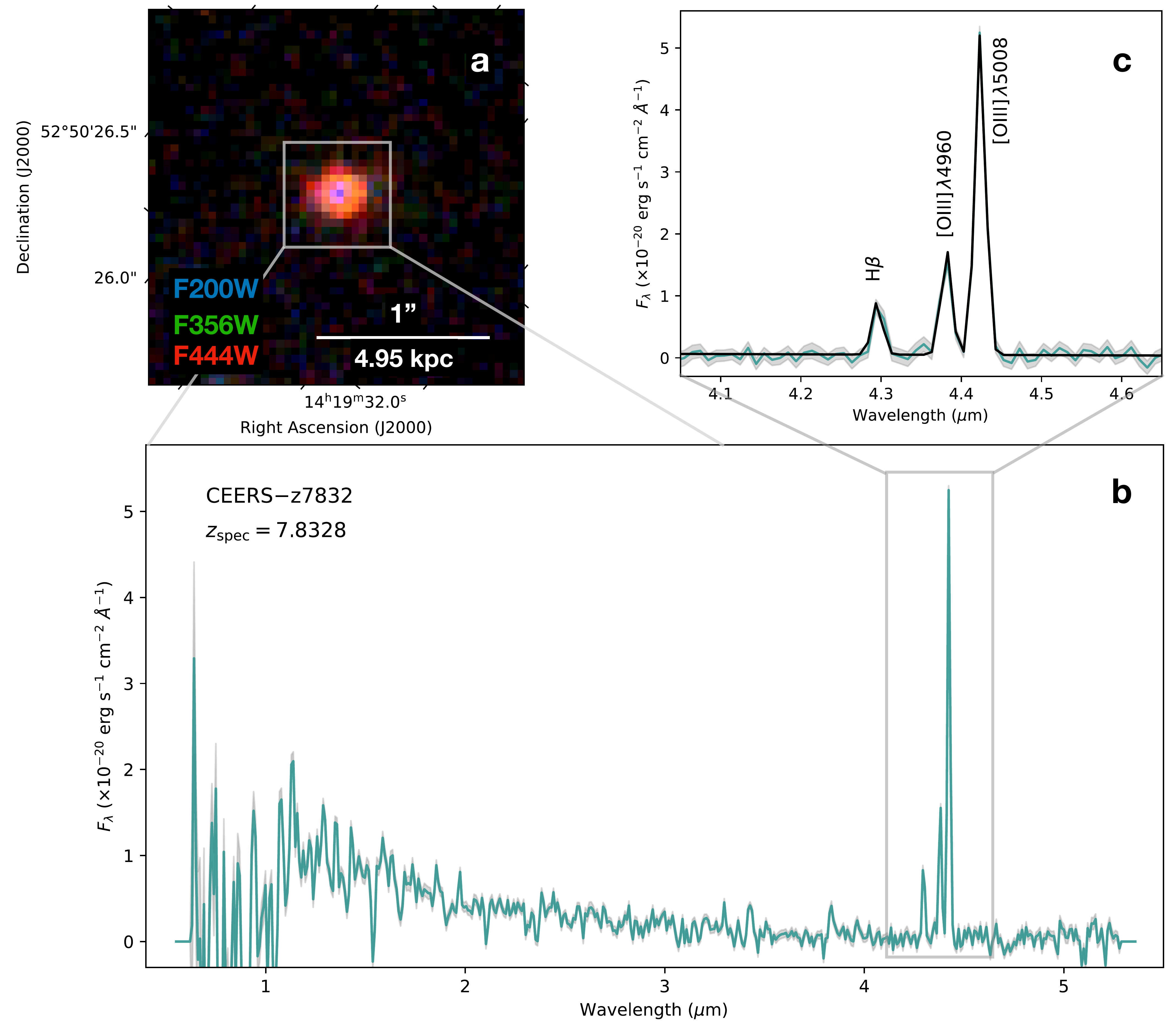}
\caption{Photometric and spectroscopic data of CEERS-$z7382$. Panel (a): False-color JWST/NIRCam RGB image zoomed in on the example galaxy (blue: F150W; green: F277W; red: F444W). The image scale and corresponding physical size at $z=7.8328$ is marked as well. Panel (b): Full NIRSpec/prism spectrum covering $0.7\mu$m to $5.2\mu$m (cyan) and associated error spectrum (grey). Panel (c): Zoom-in on the spectral region covering the nebular emission lines from the [\oiii]$\,\lambda\lambda 4960,5008$ doublet and H$\beta$. The local best-fit line and continuum model is shown by the black curve.}
\label{fig:fig1}
\end{figure}

Chemical enrichment is a key ingredient in the processes driving galaxy formation and evolution, but is also a ledger containing information on the early star-formation histories and stellar populations of the first galaxies \cite{Dayal18,Maiolino19}.
We identified galaxies at $z>7$, i.e.\ within the first 750 Myr after the Big Bang, through a blind and uniform selection of targets from recently-obtained public datasets with JWST Near-Infrared Spectrograph (NIRSpec) prism spectroscopy. These include the gravitational lensing clusters Abell 2744 (JWST DD-2756, PI: Chen) and RXJ-2129 (JWST DD-2767, PI: Kelly), and the Cosmic Evolution Early Release Science (CEERS) survey (JWST ERS-1345, PI: Finkelstein). We included all galaxies for which at least the nebular emission lines [\oiii]\,$\lambda\lambda 4959,5007$ were detected, requiring S/N$\gtrsim 3$ per wavelength element in the spectral regions covering the lines. 
Such a selection, in combination with the photometric identification of the targeted galaxies, has been shown to sample the intrinsic ultraviolet luminosity function of Lyman-break galaxies at $z=5-7$, and is therefore likely representative of the underlying high-redshift galaxy population \cite{Matthee22}.
The spectroscopic and photometric data covering these fields were obtained from the MAST archive and reduced using {\tt grizli} and custom-made pipelines (see Methods). We identify 16 targets matching the selection criteria, as summarized in Table~\ref{tab:props}. An example of the spectroscopic and imaging data obtained for each source is presented in Fig.~\ref{fig:fig1}. Throughout the paper we assume concordance $\Lambda$CDM cosmology \cite{Planck18}.

We scale the optimally-extracted 1D spectra to match the photometry for each source using a wavelength-dependent polynomial function (see Methods). This is to improve the absolute flux calibration of the spectra and correct for potential slit-losses. 
We determine the spectroscopic redshift of each galaxy by fitting a continuum model with Gaussian line profiles of the most prominent nebular emission lines imposed on to each spectrum. The redshifts span $z_{\rm spec} = 7.11 - 9.50$, corresponding to $500-750$\,Myr after the Big Bang, as summarized in Table~\ref{tab:props}. We measure line fluxes for each transition modelled with the Gaussian line profiles in the photometrically-calibrated spectra. In all cases, we detect the [\oiii]\,$\lambda\lambda 4959,5007$ doublet (by selection) and in the majority of cases we detect the Balmer line H$\beta$ and the [\oii]\,$\lambda\lambda 3726,3729$ doublet (14 and 12 out of the 16 galaxies, respectively). 
The [\oii] doublet is, however, not resolved in our spectra so we therefore only consider the sum of the doublet transitions, hence denoted [\oii]\,$\lambda 3727$. We are also not able to resolve the auroral [\oiii]\,$\lambda4363$ line transition from H$\gamma$, which is also only marginally detected in a few cases, except for the galaxy at $z=9.501$ (see Methods). This feature will generally be resolved with the NIRSpec prism only for the highest redshift sources, where the line transition is located in the redder, higher wavelength-resolution region.

We model the spectral energy distribution (SED) of each galaxy using the Bayesian Analysis of Galaxies for Physical Inference and Parameter EStimation ({\sc Bagpipes}) software package \cite{Carnall18}. We infer the physical properties by jointly modeling the spectroscopic and photometric data, covering $0.7-5.2\,\mu$m ($\approx1000-5000$\,\AA\ in the rest-frame) for each source. We assume a non-parametric star-formation history (SFH), since constant SFHs will typically underestimate the stellar mass build-up from older populations \cite{Whitler23}. Indeed, spatially resolved galaxies at $z>7$ have been observed to host older stellar populations that are otherwise outshone in integrated analyses by the younger, burstier stellar population \cite{Gimenez22}. 
We correct the SED outputs for the magnification factors due to lensing listed in Table~\ref{tab:props}, assuming the GLAFIC-v4 lens model for the Abell 2744 cluster, and a specific lens model determined for the RXJ-2129 galaxy cluster (see Methods).
We infer stellar masses in the range $M_\star = 10^{7.5}-10^{10.0}\,M_\odot$, mass-weighted ages of the stellar populations of $\tau_{\rm mass} = 200-400$\,Myr, and low dust attenuation (typically $A_V \lesssim 0.1$\,mag).
The stellar masses and mass-weighted ages are consistent with previous estimates of galaxies at $z>7$ assuming a similar flexible SFH, but are substantially larger (0.5--1\,dex) than inferred from models assuming a likely too simple constant SFH (see Methods). 
We apply an additional 0.2\,dex systematic uncertainty to the statistical errors on the derived stellar masses in the following analysis, due to the systematic uncertainties stemming from the exact choice of SED modelling technique used \cite{Pacifi22}.

\begin{figure}
\centering
\includegraphics[width=0.8\textwidth]{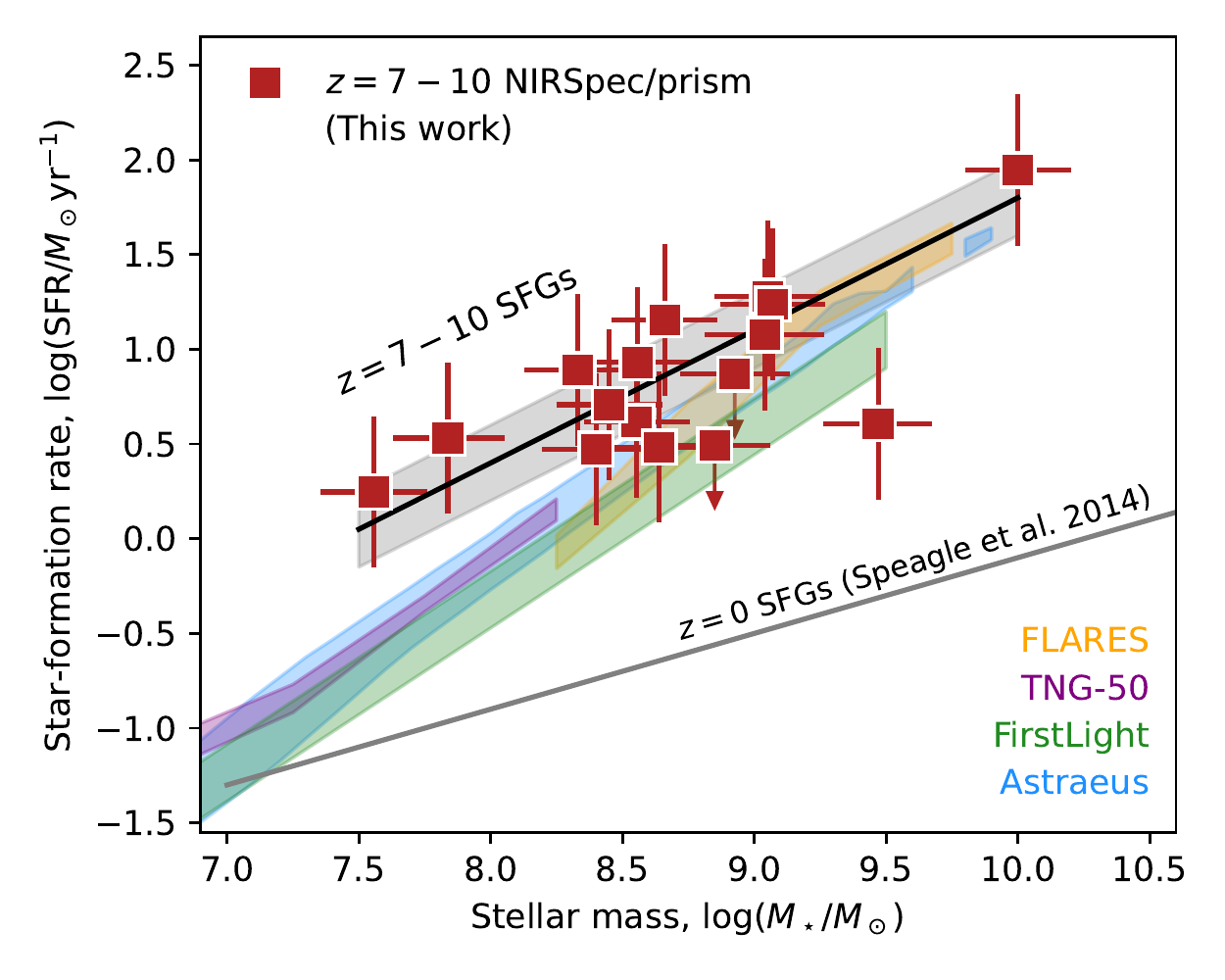}
\caption{The star-forming galaxy SFR-$M_\star$ main-sequence at $z=7-10$. Red squares show the primary sources analyzed in this work where the errorbars represent the combined statistical and systematic uncertainties. The best-fit linear relation is shown by the black curve, with the empirical scatter represented by the grey-shaded region. The colored regions represent the predicted SFR-$M_\star$ relations from $z=7$ to $z=10$, extracted from the simulations denoted in the legend. The bottom dark-grey curve shows the star-forming galaxy SFR-$M_\star$ main-sequence at $z=0$ \cite{Speagle14}.}
\label{fig:sfrmstar}
\end{figure}

We infer the star-formation rate (SFR) for each galaxy based on the H$\beta$ flux measurements \cite{Kennicutt98} as 
\begin{equation}
    {\rm SFR}_{\rm H\beta}(M_\odot/{\rm yr}) = 5.5\times 10^{-42} L_{\rm H\beta}({\rm erg/s})\times f_{\rm H\alpha/H\beta} ~ ,
\end{equation}
assuming a Kroupa initial mass function (IMF) \cite{Kroupa01} and the theoretical $f_{\rm H\alpha/H\beta} = 2.86$ ratio from the Case B recombination model at $T_e=10^{4}$\,K \cite{Osterbrock06}. We measure SFRs in the range $2-100\,M_\odot\,{\rm yr^{-1}}$, based on the photometrically-calibrated spectra and taking into account the relevant magnification factors. 
The statistical uncertainties from the H$\beta$ line flux measurements are typically $10-30\%$, whereas the uncertainties from the choice of the IMF is $\approx 20-30\%$. We therefore conservatively assume a 0.4 dex total uncertainty for each SFR measurement as summarized in Table~\ref{tab:props}. Combined with the stellar masses inferred from the photometry, we find that the identified galaxies at $z= 7-10$ all seem to follow the same SFR-$M_\star$ star-forming galaxy ``main-sequence'' (SFMS), quantified by $\log({\rm SFR}/M_\odot\,{\rm yr^{-1}}) = 0.7\times \log(M_\star / M_\odot) - 5.2$ with a scatter of 0.4 dex (see Fig.~\ref{fig:sfrmstar}). 
This is substantially higher, by $5\sigma$, than the observed local, $z\approx 0$ SFMS \cite{Speagle14}, but consistent with previous mass-complete estimates of the SFMS at $z\approx 6$ \cite{Sandles22}.

We compare our observational results to predictions of the SFMS at $z=7-10$ from recent cosmological simulations in Fig.~\ref{fig:sfrmstar}, including: Astraeus \cite{Hutter21}, the First Light and Reionisation Epoch Simulations (FLARES) \cite{DSilva22}, FirstLight \cite{Ceverino18}, and the TNG50 simulations \cite{Nelson2019TNG50}. The simulations appear to be in good agreement internally, and are overall consistent with our data. We find marginal evidence for a flatter slope of the SFMS at $z=7-10$ than predicted by simulations. However, this is mainly based on the two least massive galaxies in our sample. The simulations are able to reproduce our results for the majority the sample, at stellar masses $>10^{8}\,M_\odot$.

To determine the gas-phase metallicity through the inferred oxygen abundance, $12+\log$(O/H), we rely on the strong-line diagnostics of the most prominent nebular emission lines as we are, in most cases, not able to resolve or securely constrain the auroral [\oiii]\,$\lambda 4363$ line transition. 
We adopt the strong-line calibration based on the [\oiii]$\,\lambda 5008$/H$\beta$ line ratio (denoted R3), constrained over a large range of gas-phase metallicities, $12+\log{\rm (O/H) = 6.9 - 8.9}$ \cite{Nakajima22}. This particular line ratio was chosen due to the proximity of [\oiii]$\,\lambda 5008$ and H$\beta$ in wavelength space, such that the derived line ratios are not affected by systematic issues due to the flux calibration of the spectra or dust attenuation. Further, it has been established with a small empirical scatter (0.16\,dex), and is supported by recent theoretical considerations \cite{Nakajima22} and empirical constraints based on the detection of the auroral [\oiii]\,$\lambda 4363$ line in a sizable sample of galaxies at $z>5$ \cite{Nakajima23}. 
Specifically, we adopt the ``Large EW'' R3 calibration based on the observed equivalent width of H$\beta$, EW$_{\rm H\beta}>200$\,\AA, of the galaxies in our sample, representing the stronger ionization states. For the galaxy RXJ-$z9500$, we place lower bounds on the inferred metallicity based on the marginal detection of [\oiii]\,$\lambda 4363$ using the direct $T_e$ method (Methods), consistent with that inferred from the R3 diagnostic. For a subset (3 out of 16) of the galaxies we infer the metallicities using other calibrations, since their [\oiii]$\,\lambda 5008$/H$\beta$ line ratios were at the turnover of the R3 calibration and therefore do not allow us to robustly constrain the metallicity (see Methods).

\begin{figure}
\centering
\includegraphics[width=0.8\textwidth]{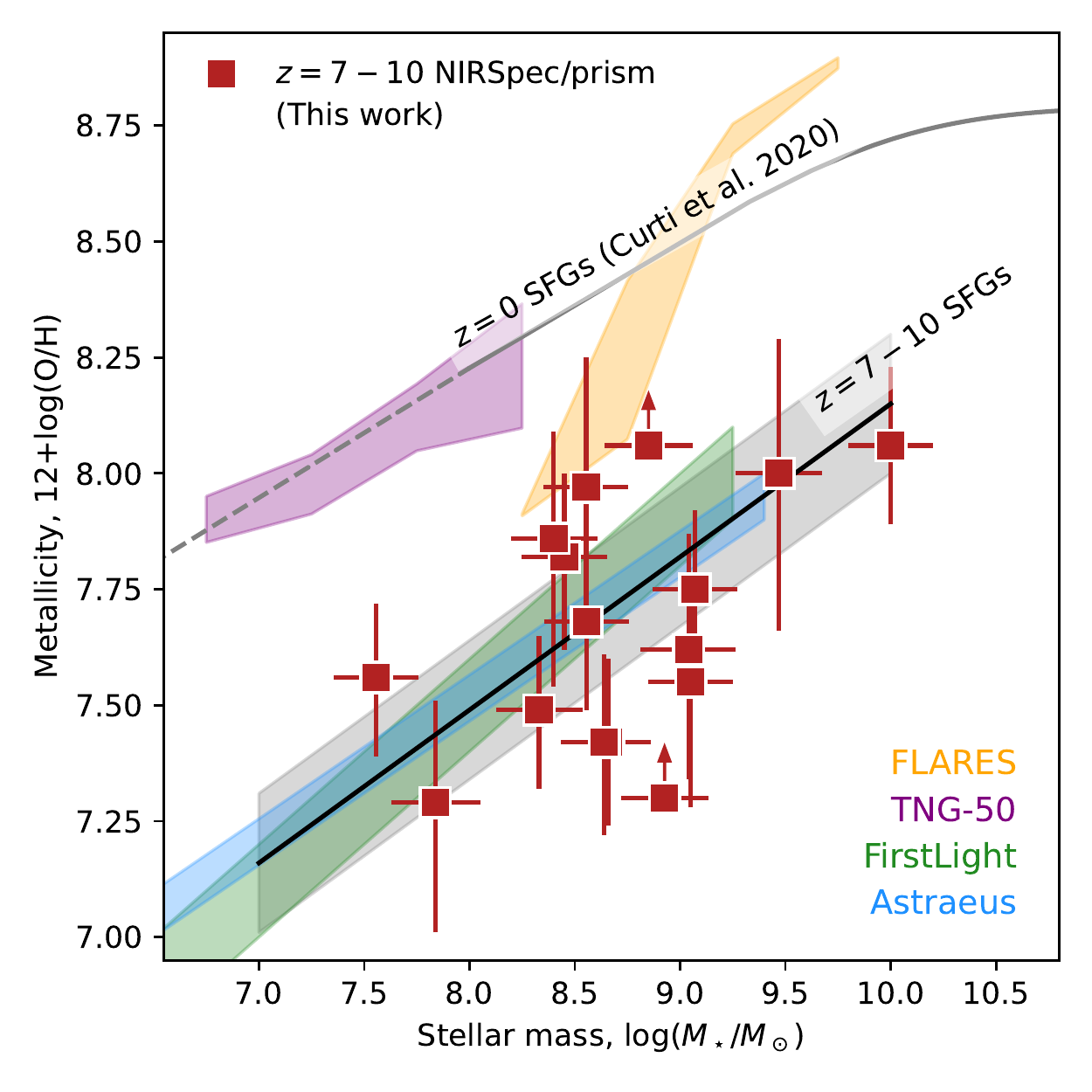}
\caption{The stellar mass-metallicity relation of galaxies at $z=7-10$. The symbol notation for our observations and the simulations are identical to Fig.~\ref{fig:sfrmstar}, except for the dark-grey dashed and continuous line which represents the mass-metallicity relation of galaxies at $z=0$ from \cite{Curti20} (extrapolated below $M_\star = 10^{8}\,M_\odot$).
}
\label{fig:massmet}
\end{figure}

The mass-metallicity relation of the galaxies at $z= 7-10$ are shown in Fig.~\ref{fig:massmet}. We observe that they all follow the same approximate log-linear relation, quantified as $12+\log$(O/H) = $0.33 \times \log(M_\star/M_\odot) + 4.85$, with a 0.3 dex scatter. 
The observed galaxies show substantially lower gas-phase metallicities, at $5\sigma$ confidence, than local star-forming galaxies at $z\approx 0$ at equivalent stellar masses \cite{Curti20}. 
We again compare our observations to predictions of the mass-metallicity relation of galaxies at $z=7-10$ from simulations, including constraints from Astraeus \cite{Ucci23}, FirstLight \cite{Langan20}, FLARES \cite{Lovell21}, and TNG50 \cite{Nelson2019TNG50}. Overall, the simulations show a substantial scatter in their predictions of the mass-metallicity relation at $z=7-10$. 
Our observations are able to put strong constraints on the stellar and chemical build-up in these various simulations, and seem to best match the prescriptions used in Astraeus and FirstLight. The slopes of the empirical relation and those predicted from simulations are generally consistent, except for FLARES which predicts a more rapid chemical enrichment with increasing stellar mass.

\begin{figure}
\centering
\includegraphics[width=0.8\textwidth]{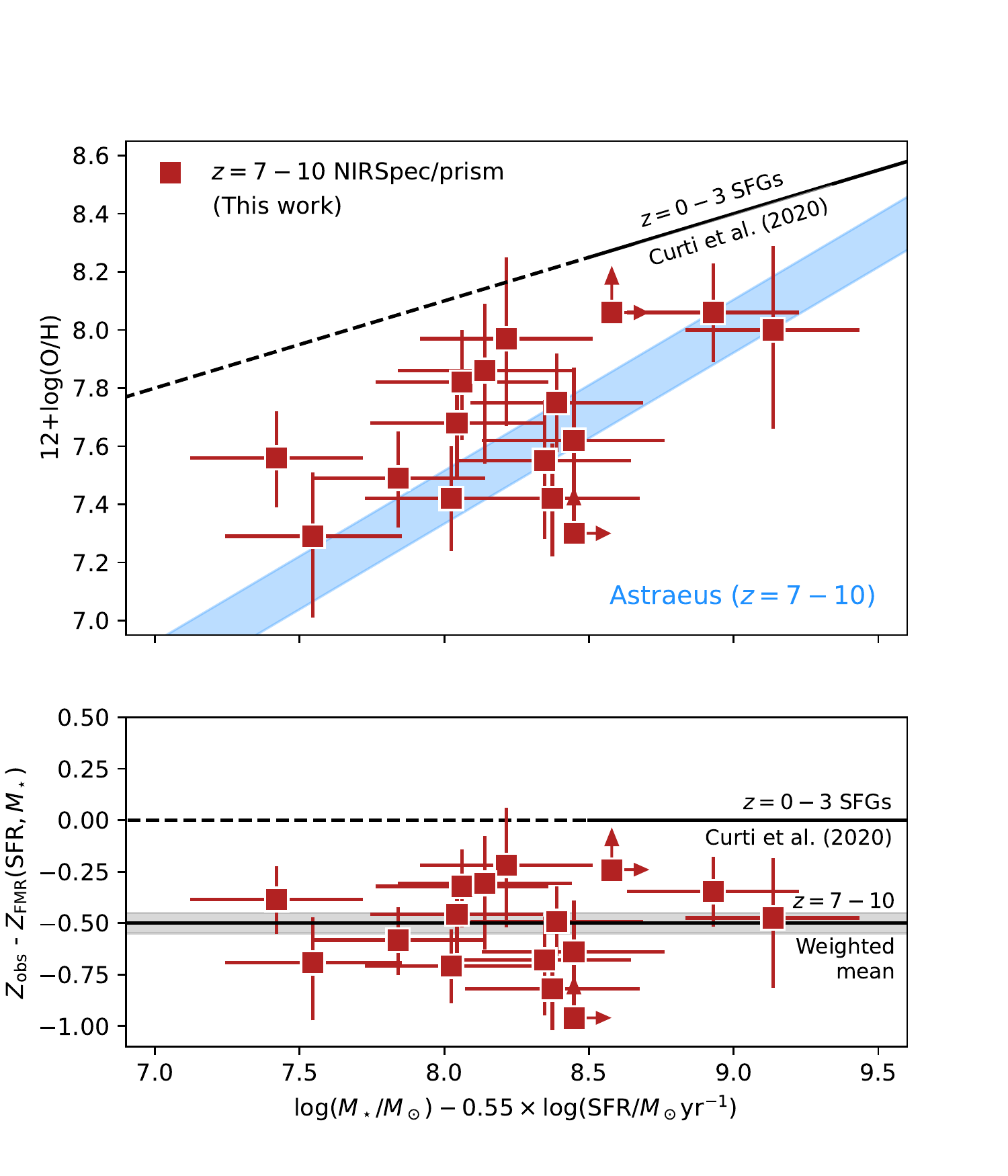}
\caption{Top panel: The fundamental-metallicity relation (FMR) of galaxies at $z=0-3$ \cite[][converted to a Kroupa IMF]{Curti20}, extrapolated below $\log(M_\star / M_\odot) - 0.55\times \log({\rm SFR}/M_\odot\,{\rm yr}^{-1}) = 8.5$. The symbol notation follows Fig.~\ref{fig:sfrmstar}, with the functional form of the high-redshift FMR predicted by the Astraeus simulations \cite{Ucci23} shown in blue. Bottom panel: Offset from the local FMR. The black solid line and grey-shaded region represent the weighted mean and standard deviation of the offset ($\Delta \log (Z/Z_\odot) = 0.50\pm 0.05$) for the primary sample of galaxies at $z=7-10$. }
\label{fig:fmr}
\end{figure}

With this unique dataset, we can for the first time test whether high-redshift galaxies follow the same fundamental mass-metallicity-SFR relation (FMR) observed for galaxies within the last 12 Gyr of cosmic time, i.e. from redshifts $z=0-3$ \cite{Mannucci10,Curti20,Sanders21}. 
Fig.~\ref{fig:fmr} shows the metallicity offsets from the local FMR based on the inferred SFR and $M_\star$ for each of the galaxies at $z= 7-10$ in our sample. We observe systematically lower metallicities ($Z_{\rm obs}$) than predicted from the FMR ($Z_{\rm FMR}$), with a weighted mean and standard deviation of $Z_{\rm obs} - Z_{\rm FMR} = -0.50\pm 0.05$ at $z= 7-10$. 
The high-redshift FMR recovered from the Astraeus simulation \cite{Ucci23} is generally consistent with our observations and predict an even larger offset from the local FMR for galaxies at $z=7-10$ at low stellar masses or comparably higher SFRs (see Fig.~\ref{fig:fmr}).
There have already been some early results hinting that the local FMR relation may not hold at redshifts $z\gtrsim 3.5$ \cite{Mannucci10,Troncoso14,Onodera16}. Most of these observations relied on limited sample sizes, however, and the majority of sources were starburst galaxies, making them overall unrepresentative of the typical galaxy population at these redshifts. The first JWST study of the three galaxies at $z\gtrsim 7$ observed as part of the ERO was also inconclusive \cite{Curti23}. The results presented here are thus the first to robustly verify this offset and the divergence from the local FMR at early cosmic times. 

The drastic drop in metallicity in these early universe galaxies could be explained by significant accretion from the intergalactic medium that rapidly replenishes the galaxies with continuous infall of neutral, pristine gas. This process effectively dilutes the metals, reducing the overall chemical abundance \cite{Maiolino19}. Gas accretion from the intergalactic medium is expected theoretically to be an important process for feeding star-forming galaxies in the first $\approx 1-2$\,Gyr of cosmic time~\cite{Lagos18}. Recent observations of abundant gas reservoirs in galaxies at $z>6$ \cite{Heintz22} support this hypothesis. In this scenario, it is the exhaustion of these pristine, intergalactic gas reservoirs that is responsible for the peak and rapid drop in the cosmic star-formation rate observed at redshift $z\approx 1–3$ \cite{Madau14}.

\clearpage
\newpage

\begin{table}[h]
\begin{minipage}{400pt}
\caption{Physical properties of the primary sample galaxies at $z= 7-10$.}\label{tab1}%
\begin{tabular}{@{}lcccccrc@{}}
\toprule
Galaxy ID  & $z_{\rm spec}$ & SFR$_{\rm H\beta}$ & $\log M_\star$ & $A_V$ & $\tau_{\rm mass}$ & $12+\log$(O/H) & $\mu$ \\
&& ($M_\odot ~{\rm yr}^{-1}$) & ($M_\odot$) & (mag) & (Myr) & & \\
\midrule
RXJ-z9500 & 9.5008 & $33.9^{+51.3}_{-20.4}$ & $8.84^{+0.02}_{-0.03}$ & $0.02^{+0.02}_{-0.01}$ & $221^{+13}_{-13}$ & $7.56^{+0.16}_{-0.17}$ (R3) & $19.2 \pm   3.6$ \\
RXJ-z8149 & 8.1496 & $7.6^{+11.6}_{-4.6}$ & $8.19^{+0.08}_{-0.06}$ & $0.05^{+0.06}_{-0.04}$ & $262^{+17}_{-19}$ & $7.29^{+0.22}_{-0.28}$ (R3)  & $2.25  \pm  0.14$ \\
\vspace{0.1cm}
RXJ-z8152 & 8.1523 & $12.4^{+18.8}_{-7.5}$ & $8.72^{+0.02}_{-0.03}$ & $0.04^{+0.03}_{-0.02}$ & $266^{+16}_{-17}$ & $7.68^{+0.18}_{-0.19}$ (R3)  & $1.46 \pm   0.03$ \\
Abell-z7878 & 7.8783 & $<9.8$     &            $9.05^{+0.06}_{-0.05}$ & $0.25^{+0.05}_{-0.04}$ &$270^{+17}_{-19}$ & $>7.30$ (R3)  & $1.33  \pm  0.04$ \\
\vspace{0.1cm}
Abell-z7885 & 7.8854 & $8.8^{+13.3}_{-5.3}$ & $8.88^{+0.02}_{-0.02}$ & $0.10^{+0.02}_{-0.02}$ & $285^{+17}_{-19}$ & $7.97^{+0.28}_{-0.30}$ (O32)  & $2.12 \pm   0.06$ \\
CEERS-z7105 & 7.1051 & $5.1^{+7.7}_{-3.1}$ & $8.45^{+0.03}_{-0.02}$ & $0.02^{+0.02}_{-0.01}$ & $304^{+22}_{-21}$ & $7.82^{+0.18}_{-0.20}$ (R3) & $--$ \\
CEERS-z7179 & 7.1792 & $7.8^{+11.8}_{-4.7}$ & $8.33^{+0.05}_{-0.03}$ & $0.06^{+0.03}_{-0.02}$ & $298^{+19}_{-21}$ & $7.49^{+0.16}_{-0.17}$ (R3) & $--$ \\
CEERS-z7175 & 7.1751 & $14.3^{+21.6}_{-8.6}$ & $8.66^{+0.02}_{-0.02}$ & $0.02^{+0.01}_{-0.01}$ & $305^{+20}_{-20}$ & $7.42^{+0.18}_{-0.18}$ (R3) & $--$ \\
CEERS-z7167 & 7.1683 & $3.0^{+4.5}_{-1.8}$ & $8.40^{+0.05}_{-0.04}$ & $0.03^{+0.02}_{-0.01}$ & $296^{+22}_{-19}$ & $7.86^{+0.23}_{-0.32}$ (R3) & $--$ \\
CEERS-z8684 & 8.6849 & $88.4^{+133.6}_{-53.2}$ & $10.0^{+0.01}_{-0.01}$ & $0.51^{+0.01}_{-0.01}$ & $398^{+2}_{-3}$ & $8.06^{+0.17}_{-0.17}$ (R3) & $--$ \\
CEERS-z7789 & 7.7805 & $<3.1$      &           $8.85^{+0.07}_{-0.06}$ & $0.04^{+0.02}_{-0.03}$ & $271^{+18}_{-17}$ & $>8.06$ (R3) &  $--$ \\
CEERS-z8718 & 8.7182 & $19.0^{+28.7}_{-11.4}$ & $9.05^{+0.03}_{-0.02}$ & $0.02^{+0.01}_{-0.01}$ & $242^{+16}_{-15}$ & $7.55^{+0.21}_{-0.27}$ (R3)  &  $--$ \\
CEERS-z7832 & 7.8328 & $17.3^{+26.2}_{-10.4}$ & $9.07^{+0.01}_{-0.01}$ & $0.10^{+0.02}_{-0.02}$ & $447^{+6}_{-9}$ & $7.75^{+0.17}_{-0.17}$ (R3) & $--$ \\
CEERS-z8612 & 8.6142 & $4.1^{+6.1}_{-2.4}$ & $9.47^{+0.04}_{-0.06}$ & $0.09^{+0.03}_{-0.02}$ & $234^{+20}_{-26}$ & $8.00^{+0.29}_{-0.34}$ (O32) &   $--$ \\
CEERS-z8172 & 8.1724 & $11.9^{+18.0}_{-7.2}$ & $9.04^{+0.10}_{-0.11}$ & $0.19^{+0.06}_{-0.06}$ & $261^{+18}_{-16}$ & $7.62^{+0.25}_{-0.28}$ (R2)  & $--$ \\
CEERS-z7453 & 7.4536 & $3.0^{+4.6}_{-1.8}$ & $8.64^{+0.05}_{-0.05}$ & $0.02^{+0.02}_{-0.02}$ & $284^{+17}_{-19}$ & $7.42^{+0.19}_{-0.20}$ (R3) & $--$ \\
\botrule
\end{tabular}
\footnotetext{{\bf Notes.} The listed properties are not corrected for the magnification factor provided in Col. 8.}
\label{tab:props}
\end{minipage}
\end{table}

\clearpage
\newpage

\backmatter

\bmhead{Acknowledgments}
We thank the reviewers for thorough and constructive reports, greatly improving the analysis and results presented in this work. We would also like to thank Daniel Ceverino for his insight and suggestions on interpreting the FirstLight simulations output. We further acknowledge the GLASS, CEERS, and UNCOVER collaborations, and are grateful that they made their early data publicly available.
K.E.H. acknowledges support from the Carlsberg Foundation Reintegration Fellowship Grant CF21-0103. C.A.M and A.H. acknowledge support by the VILLUM FONDEN under grant 37459.
The Cosmic Dawn Center (DAWN) is funded by the Danish National Research Foundation under grant No. 140. 
This work is based on observations made with the NASA/ESA/CSA James Webb Space Telescope. The data were obtained from the Mikulski Archive for Space Telescopes at the Space Telescope Science Institute, which is operated by the Association of Universities for Research in Astronomy, Inc., under NASA contract NAS 5-03127 for JWST. 

\bmhead{Author contributions}

K.E.H. wrote the manuscript and led the analysis. G.B. reduced and extracted the photometric and spectroscopic data. C.G.-A. and V.B.S. performed the SED modelling. All authors contributed to the manuscript and aided the analyses and interpretations.

\bmhead{Declarations}

The authors declare no competing financial interests. 

\bmhead{Correspondence}

Correspondence and requests for materials should be addressed to K.E.H. (email: \url{keheintz@nbi.ku.dk}).

\clearpage
\newpage

\section*{Methods}

\bmhead{Cosmology}

Throughout this work, we assume concordance, flat $\Lambda$CDM cosmology with $\Omega_{\rm m} = 0.315$, $\Omega_{\Lambda} = 0.685$, and $H_0 = 67.8$\,km\,s$^{-1}$\,Mpc$^{-1}$ \cite{Planck18}. We use the cosmology distance calculator from \texttt{astropy} \cite{astropy} to infer the luminosity distances $D_L$ to the given redshifts.

\bmhead{Observations and data reduction}

This work is primarily based on the photometric and spectroscopic data taken as part of the public UNCOVER (GO-2561), GLASS (ERS-1324), and CEERS (ERS-1345) surveys, and including observations obtained through the JWST Director's Discretionary time (DD-2756, PI: Chen and DD-2767, PI: Kelly). The NIRCam observations were typically taken in six broad-band filters (F115W, F150W, F200W, F277W, F356W, F444W). The photometric data are extracted from the catalog presented by Brammer et al. (in prep), which provides a compilation of the JWST ERS photometric data released to date. The raw data entering this catalog has been reduced using the public software package \texttt{grizli} \cite{Brammer19}, which masks imaging artifacts, provides astrometric calibrations based on the Gaia DR3 catalog, and shifts the images to a common pixel scale of $0.''04$/pixel using \texttt{astrodrizzle}. This catalog also includes the most recent updated photometric zero-points, and are all corrected for the Milky Way extinction. The catalog is published at this MAST DOI: \texttt{10.17909/g2av-f122}. We use the aperture-matched photometry derived using a diameter of $0.''5$ from the catalog. A subset of the galaxies in our sample with NIRSpec prism spectra from the CEERS survey (CEERS-$z7789$, -$z8612$, -$z8172$, and -$z7453$) are not covered by the JWST NIRCam mosaic. For those targets we use archival data from the {\em Hubble Space Telescope} (HST) processed in the same manner as the JWST mosaics described above, and published at this MAST DOI: \texttt{10.17909/qhte-a822}.

\begin{figure}
\centering
\includegraphics[width=0.8\textwidth]{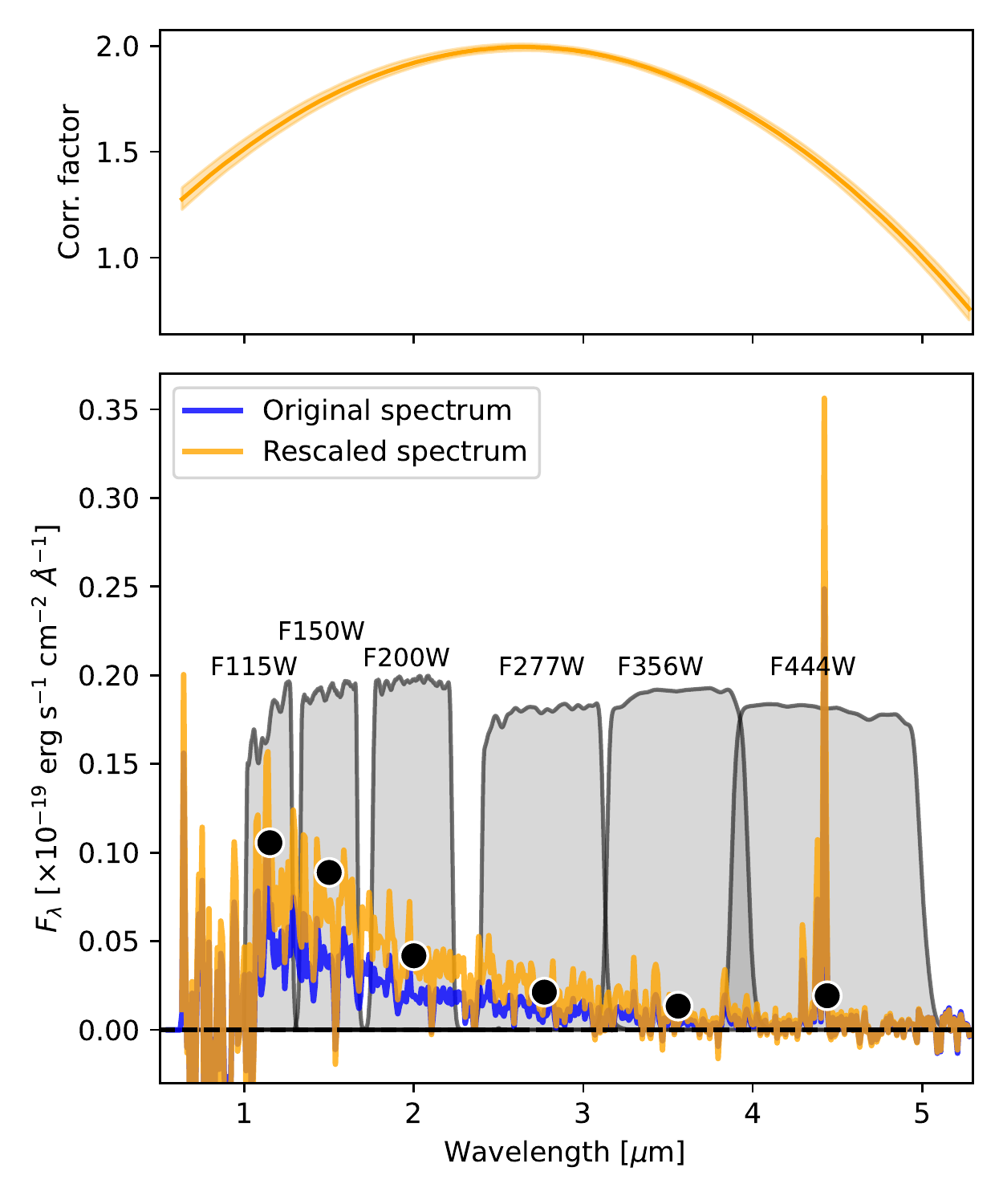}
\caption{The original (blue) and rescaled (orange) 1D spectrum of the galaxy CEERS-$z7832$. The NIRSpec prism spectrum has been scaled to match the NIRCam photometry to improve the accuracy of the flux calibration and account for potential slit-losses. The NIRCam filter curves used to photometrically calibrate the spectrum is shown by the grey passbands. The derived polynomial function, which we apply to correct the original spectrum, is shown in the top panel. }
\label{efig:scalespec}
\end{figure}

The 1D spectra are extracted from a custom-made pipeline \cite{msaexp}, that utilizes the Stage 2 output from the MAST archive. This code performs the standard wavelength, flat-field and photometric calibrations to the individual NIRSpec exposure files\footnote{i.e., calibration levels 1 and 2 with \texttt{jwst} version 1.8.2}. We use the JWST reference files associated with the CRDS context \texttt{jwst\_1027.pmap}. We generate the full combined 2D spectra and optimal \cite{Horne86} 1D extractions with scripts that extend the standard JWST pipeline functionality \cite{msaexp}. We further scale the overall flux densities of the spectra to the derived photometry, matching the integrated flux within the available passbands using a wavelength-dependent polynomial function. This is to improve the absolute flux calibration of the spectra and take into account potential slit-losses, and is further used in the SED modelling of each source. An example is shown in Fig.~\ref{efig:scalespec} for the galaxy CEERS-$z7832$. The required scaling factors are typically of the order $1.2-2$ and varies by 20-30\% across the spectrum. The NIRSpec prism spectra typically cover the full wavelength range from $0.7\mu$m to $5.2\mu$m, except for RXJ-$z8149$ and CEERS-$z7179$, which have gaps at $<3\mu$m and $1.5-4\mu$m, respectively. The spectral resolution range from $\mathcal{R} = 50$ in the blue end to $\mathcal{R} = 400$ in the red end \cite{Jakobsen22}. The full set of reduced and photometrically-calibrated 1D NIRSpec/prism spectra are shown in Fig.~\ref{efig:allspec}.

\bmhead{Lens models}

To infer the relevant magnification factors for each source in the RXJ-2129 and Abell-2744 fields, we adopt the lens model presented by \cite{Zitrin15} for the RXJ-2129 galaxy cluster and consider the lens models compiled for the Hubble Frontier fields \footnote{\url{https://archive.stsci.edu/prepds/frontier/lensmodels/}} for the Abell 2744 cluster. We adopt the GLAFIC-v4 lens model results in this work, but overall find that magnification factors are consistent within $5-10\%$ for the various online lens models. The magnification factors are summarized in Table~\ref{tab:props}, typically of the order $\mu = 1-2$ except for the galaxy RXJ-$z9500$ which has $\mu = 19.2\pm 3.6$ \cite{Williams22}. 

\bmhead{SED modelling}

The SEDs of each galaxy have all been modelled using {\sc Bagpipes} \cite{Carnall18} to infer their physical properties such as the stellar masses and mass-weighted ages. This code uses the stellar population models from Bruzual \& Charlot \cite{Bruzual03}. We fix the redshifts to $z_{\rm spec}$ and include nebular emission via {\sc Cloudy} \cite{Ferland17}. 
We jointly model the photometric and spectroscopic data for each source, which greatly improves the statistical uncertainties from the posterior distributions. We further modify the default capabilities of {\sc Bagpipes} to incorporate a wavelength-dependent spectral resolution matching the NIRSpec prism spectra.
We allow the ionization parameter, $U$, to vary between $-3 < \log (U) < -1$, to account for the typically higher ionization parameters at $z\gtrsim 6$ \cite{Sugahara22}. We assume the attenuation curve from Salim et al. \cite{Salim18}, which better represent low-mass galaxies, allowing the slope ($\delta$) to vary from $-0.9 < \delta < 0.1$ and the bump strength ($B$) from $0 < B < 3$. We assume Gaussian priors on the various dust parameters and infer $A_V\lesssim 0.1$\,mag for the majority of galaxies. We assume a Kroupa 2001 initial mass function (IMF) \cite{Kroupa01} and set the metallicity to vary between $0 < Z/Z_\odot < 0.5$. 

Based on the accurate spectro-photometric modelling, we incorporate a more flexible, non-parametric SFH for each galaxy \cite{Iyer19}. We find that this effectively increases the inferred stellar masses and mass-weighted ages of the stellar population by 0.5-1 dex compared to more simple, constant SFHs, as also demonstrated in previous work \cite{Leja19,Tacchella22,Carnall23,Whitler23}. This model fits for quartiles of total mass formed, with a prior included to account for possible hidden mass due to ``outshining" from the younger, burstier star-forming populations and includes an additional prolonged, less active period of star formation. Physically, this corresponds to including the older, more massive stellar populations recovered from spatially-resolved analyses \cite{Gimenez22}.
Table~\ref{tab:props} reports the median and 16th and 84th percentiles from the resulting posterior distributions on the stellar mass, dust attenuation, and mass-weighted ages of the stellar populations for each galaxy. 

We tried various sets of combinations using different parametrizations of the SFH and dust laws, including a constant and delayed SFH and a Salim \cite{Salim18} vs. Calzetti \cite{Calzetti} dust attenuation curve. An exponentially declining SFH is likely not an accurate representation of these very high-redshift galaxies and may overestimate the inferred stellar masses with increasing SFRs. Moreover, the constant SFH is likely a too simple representation of the more complex SFHs expected for these galaxies, as predicted from simulations \cite{Ma15}. Assuming a Salim over a Calzetti-like attenuation curve effectively decreases the inferred $A_V$ due to the steeper functional form of the Salim curve. The slope of the attenuation curve is related to the dust opacity and grain size distribution, which introduces and apparent dependence on the stellar mass as well \cite{Salim18} such that more more massive galaxies have shallower slopes. However, we found that our results were remarkably consistent overall (within 0.1 dex in stellar mass estimates), independent of the exact dust prescriptions used.

\bmhead{Line flux measurements}

To estimate the line fluxes for each galaxy, we model the most prominent redshifted nebular emission lines: H$\beta$, H$\gamma$, [\oiii]\,$\lambda 4959,5007$, and [\oii]\,$\lambda 3727$ imposed on the photometrically-calibrated spectra with simple Gaussian line profiles. The spectral resolution is insufficient to determine more accurate line-profile kinematics. The redshift $z_{\rm spec}$ and line widths are tied for all the lines, such that only the line fluxes can vary across transitions. We further fix the line flux ratio of the [\oiii]$\,\lambda\lambda 4960,5008$ doublet to the theoretical predicted value of $f_{\rm [OIII]-4960} / f_{\rm [OIII]-5008} = 1/3$ in the fits, based on the relative transition probabilities from the excited $^1D_2$ state to $^3P_1$ and $^3P_2$, respectively \cite{Osterbrock06}.

\bmhead{Metallicity calibrations}

To determine the gas-phase metallicity for each galaxy we use the strong-line diagnostics from Nakajima et al. \cite{Nakajima22}, since we are not able to resolve the auroral [\oiii]\,$\lambda 4363$ line transition in all but one case. Instead, we chose the R3 diagnostic since it has a small empirical scatter (0.16 dex), and is supported by recent theoretical framework \cite{Nakajima22} and empirical constraints based on direct metallicities estimates through the $T_e$-method \cite{Nakajima23}. Further, the [\oiii]$\,\lambda 5008$ and H$\beta$ emission lines are closely separated in wavelength space and this calibration therefore introduce less uncertainties due to the flux calibration and dust attenuation of the spectrum. Specifically, we adopt the ``Large EW'' calibration, representing the stronger ionization states of high-redshift galaxies \cite{Nakajima23}, and also consistent with the inferred large H$\beta$ EWs $>200\,\AA$ for the galaxies in our sample. The inferred metallicities are summarized in Table~\ref{tab:props}.
Comparing the estimates from the R3 calibrations to the R2, R23, and O32 calibrations from \cite{Nakajima22}, all requiring the additional detection of [\oii]$\,\lambda 3727$, we find that they are overall consistent within 0.1-0.2 dex. 
For three of the galaxies in our sample, Abell2744-z7885, CEERS-z8612, and CEERS-z8172, the metallicity could not be constrained using the R3 diagnostic since their line ratios were at the turnover of this particular calibration. Instead we use a combination of the R2 and O32 calibrations since the derived line fluxes allow for more accurate metallicity estimates due to the distinct functional form of these diagnostics. 


For the one galaxy RXJ-$z9500$, first presented by Williams et al. \cite{Williams22}, we constrain the metallicity based on the marginal detection of the auroral [\oiii]\,$\lambda 4363$ line at $f_{\rm [OIII]-4363} \lesssim 5.0\times 10^{-19}$\,erg\,s$^{-1}$\,cm$^{-2}$, see Fig.~\ref{efig:auroralmet}. We follow the iterations outlined by Izotov et al. \cite{Izotov06} and solve for the oxygen abundance as a function of electron temperature and density. We derive electron temperatures $T_e \lesssim 2.8\times 10^{4}$\,K and gas-phase metallicities $12+\log$(O/H) $\gtrsim 7.2$ for all $n_e < 10^{4}$\,cm$^{-3}$. This is consistent with the inferred estimate from the R3 calibration of $12+\log$(O/H) $= 7.56^{+0.16}_{-0.17}$ and with previous results from Williams et al. \cite{Williams22}.

\bmhead{Simulations of the mass-metallicity relation at $z=7-10$}
In the Main Text, we found a large scatter in the predictions for the mass-metallicity relation of galaxies at $z=7-10$ across a suite of state-of-the-art cosmological simulations.
The predictions from the TNG50 simulations \cite[extracted using the abundance ratios of gas cells weighted by the star formation rate within twice the half mass radius of galaxies; ][]{Nelson2019TNG50,Pillepich2019TNG50} at $z=7-10$ almost directly follow the observed mass-metallicity relation for galaxies at $z=0$ \cite{Curti20}, which is inconsistent with the overall evolution of this relation, even at $z< 3$. The FLARES simulations \cite[measured by using the abundance ratios of gas particles weighted by their star formation rate within a 30 pkpc radius from the centre of potential;][]{Lovell21,Vijayan21}, when extrapolated to low stellar masses, seem to be marginally consistent with the observed mass-metallicity relation for galaxies at $z= 7-10$. However, the slope appears to be significantly steeper than observed, effectively predicting more metal-enriched galaxies at $M_\star \gtrsim 10^{8}\,M_\odot$. It should be noted, however, that both the absolute normalization and the slope of the mass-metallicity relation are subject to the uncertainties in the metal yields of stars of all masses as well as the stellar feedback models implemented in the different simulations. For this reason, predicting the relative evolution of the mass-metallicity relation with redshift has been the most important so far. We also note that previous works, when comparing to theoretical predictions \cite[e.g.][]{Curti23}, assume that 35\% of the total metal fraction is in oxygen when deriving the oxygen abundance (O/H), which will likely be different at these redshifts due to $\alpha-$enhancement \cite{Wilkins22}.

The Astraeus \cite{Ucci23} and FirstLight \cite{Langan20} simulations appear to accurately match our empirical results. 
In the Astraeus simulation, masses of the metal and gas reservoirs are evolved by applying the semi-analytic galaxy evolution model described in \cite{Hutter21} and \cite{Ucci23} to the merger trees of halos identified in the dark matter-only {\it Very Small MultiDark Planck} N-body simulation. Gas-phase metallicities are then derived by taking the ratio between the oxygen and gas mass reservoirs, assuming 
the latest stellar metal yields from \cite{Kobayashi20}.
The FirstLight simulations output are taken directly from the literature \cite{Langan20}, but corrected by 0.21 dex to match the most recent estimate of the solar metallicity, $12+\log{\rm (O/H)_\odot}=8.69$ \cite{Asplund09} (from the assumed $12+\log{\rm (O/H)_\odot}=8.9$).

\bmhead{Robustness of the high-redshift FMR offset} 

To compare the local FMR parametrized by Curti et al. \cite{Curti20} to the high-redshift observations, we converted their results based on the Chabrier \cite{Chabrier03} IMF to match the Kroupa IMF \cite{Kroupa01} assumed in this work. Effectively, this increases the stellar masses by $\approx 10\%$ and decreases the SFRs by $\approx 20\%$. Since the functional form of the metallicity as a function of the stellar mass and SFR is prescribed as $\log (M_\star/M_\odot)-0.55\times \log({\rm SFR}/M_\odot {\rm yr}^{-1})$ in the FMR, this change in IMF has a negligible effect. Even if the uncertainties associated with the metallicity predictions provided by the FMR parametrization in \cite{Curti20} increases to $\approx 0.3$\,dex in the low-mass–high-SFR regime we are probing with this high-redshift galaxy sample, we still observe a systematic offset at $2\sigma$ confidence. However, based on the high-redshift FMR predicted by the Astraeus simulations \cite{Ucci23}, the offset might increase even more at the low-mass–high-SFR regime, substantiating our results even more significantly. 

As detailed above, assuming a simple constant SFH in the SED modelling of each galaxy will yield lower stellar mass estimates by 0.5-1 dex. Based on these estimates, which are effectively lower bounds, we still observe a factor of $\approx 2$ offset in the high-redshift galaxy sample from the local FMR at $2\sigma$ confidence. Accounting for a potential lower H$\alpha$-to-SFR calibration due to the higher efficiency of ionizing photon production in more metal-poor environments \cite{Shapley23}, will systematically decrease the quantity $\log (M_\star/M_\odot)-0.55\times \log({\rm SFR}/M_\odot {\rm yr}^{-1})$ by 0.2 dex. Even with these more conservative estimates we still recover a significant offset from local FMR in the metallicities of the high-redshift galaxy sample. 

Sanders et al. \cite{Sanders21} established that the local FMR \cite{Mannucci10,Curti20} holds at $z\approx 2.3$ and $z\approx 3.3$, based on an extensive sample of galaxies from the MOSDEF survey. This suggests that galaxies evolve through a smooth baryonic growth process in a state of near-equilibrium from redshifts $z=0-3.3$, i.e. from 2 Gyr after the Big Bang to the present. Comparing our observations to the recent sample of strong [\oiii]-emitting galaxies at $z= 5-7$ from Matthee et al. \cite{Matthee22} in Fig.~\ref{efig:fmrcomp} reveal no strong evolution from $z\approx 6$ to $z\approx 8$. The equilibrium state of galaxy growth and evolution thus appears to already be broken at $z\approx 6$, i.e. at 1 Gyr after the Big Bang. In Fig.~\ref{efig:fmrcomp}, we also compare our observations to the three galaxies at $z>7$ identified in the Early Science Observations (ERO), with inferred physical properties from \cite{Curti23}. These three sources all have metallicities derived from the direct $T_e$-method, and are overall well in agreement with our results. The independent measurements at $z=5-7$, and through the direct $T_e$-method for the few galaxies at $z>7$, thus greatly substantiate and support our results. Further, they imply that galaxies down to at least $z\approx 6$ are still subject to significant gas accretion from the intergalactic medium, causing a dilution of the chemical abundances in the interstellar medium of these sources. This is supported by the abundant neutral atomic gas reservoirs observed in galaxies at this epoch \cite{Heintz22,Heintz23}. The pristine, intergalactic gas is likely exhausted by $z\approx 3$, about 1 Gyr later, at which point the local FMR is established.

\clearpage
\newpage

 
\begin{figure}
\centering
\begin{subfigure}[t]{0.45\textwidth}
\centering
\includegraphics[width=\textwidth]{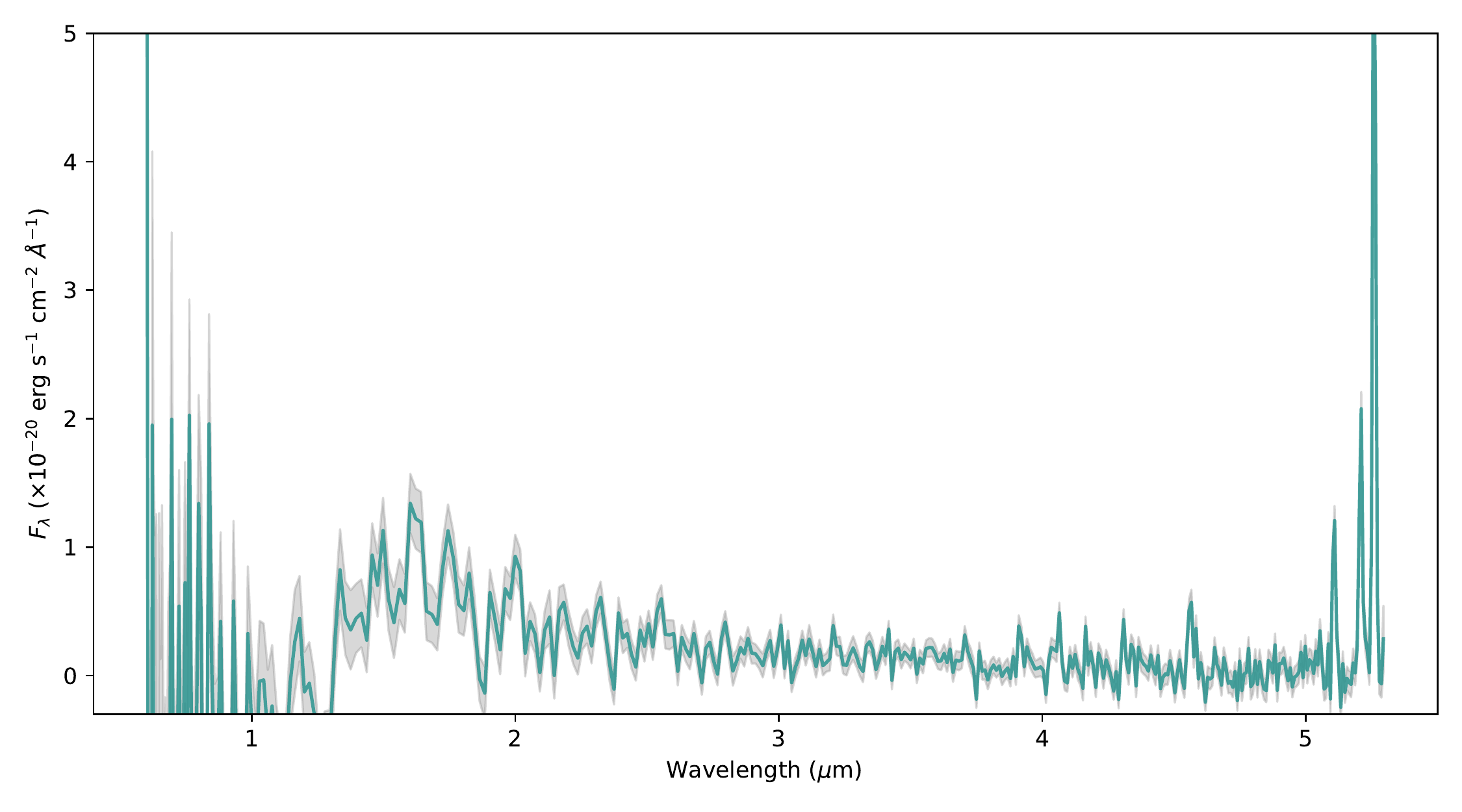} 
\caption{RXJ-$z9500$} 
\end{subfigure}
\hfill
\begin{subfigure}[t]{0.45\textwidth}
\centering
\includegraphics[width=\textwidth]{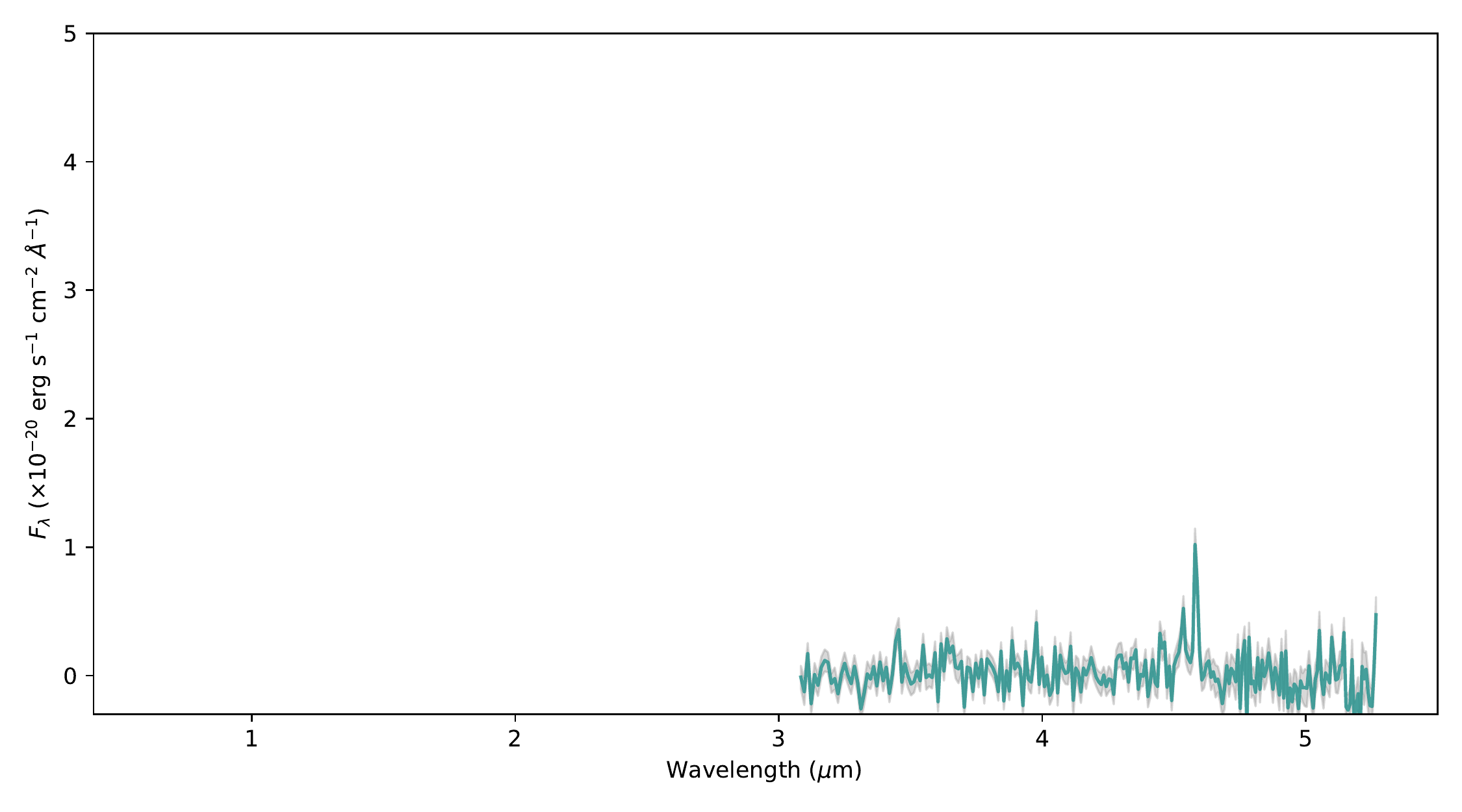} 
\caption{RXJ-$z8149$} 
\end{subfigure}
\hfill
\begin{subfigure}[t]{0.45\textwidth}
\centering
\includegraphics[width=\textwidth]{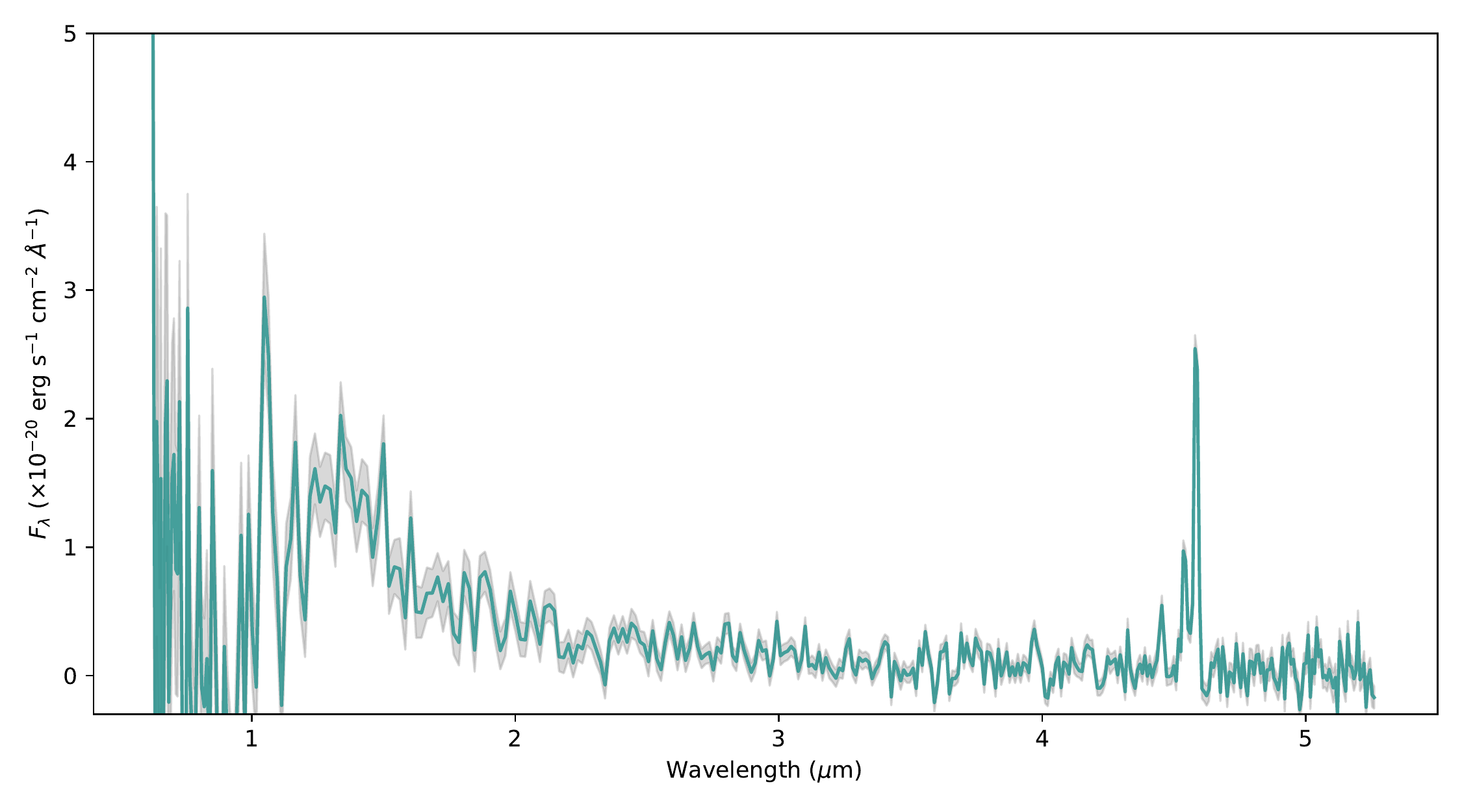} 
\caption{RXJ-$z8152$} 
 \end{subfigure}
 \hfill
\begin{subfigure}[t]{0.45\textwidth}
\centering
\includegraphics[width=\textwidth]{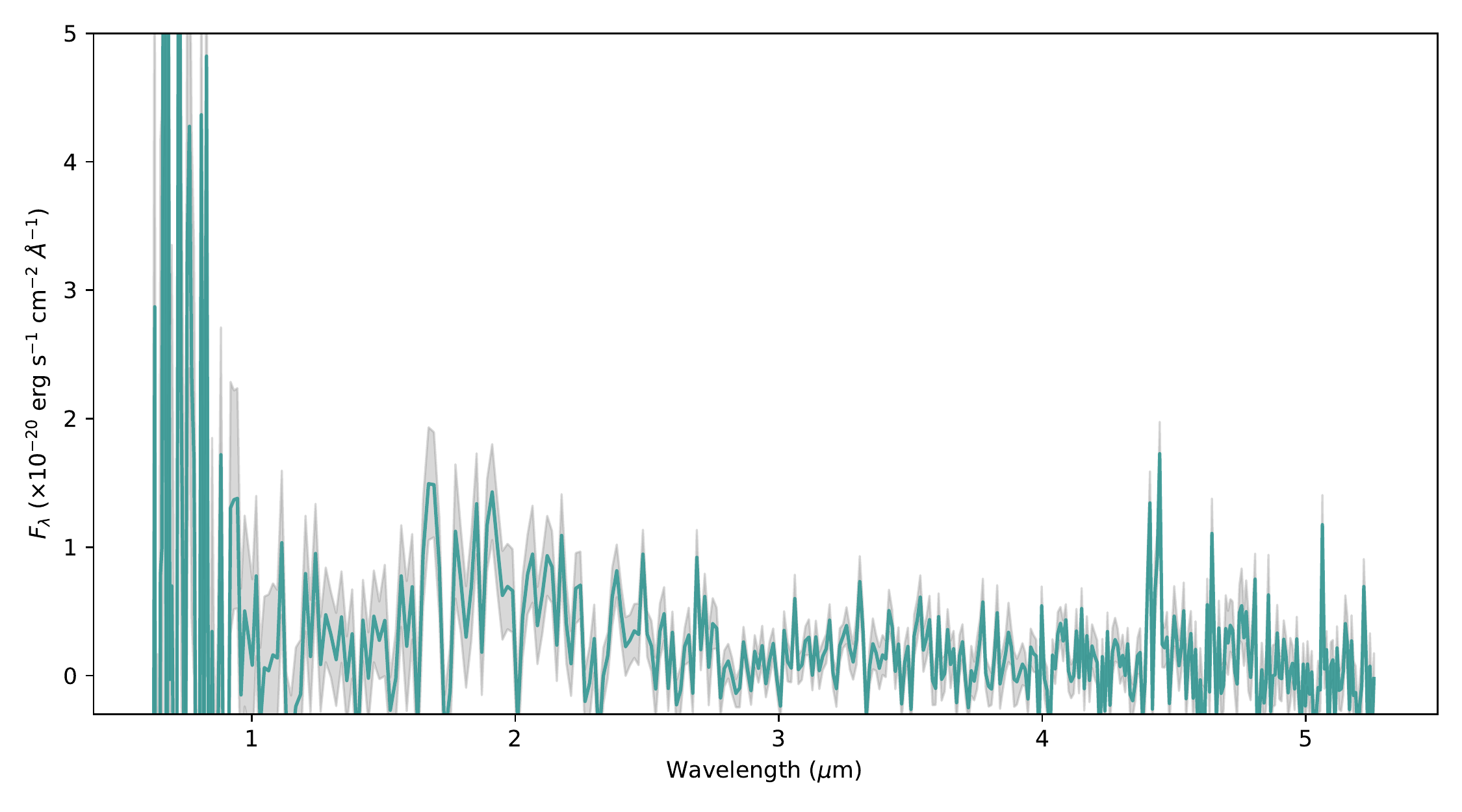} 
\caption{Abell-$z7878$} 
 \end{subfigure}
 \hfill
\begin{subfigure}[t]{0.45\textwidth}
\centering
\includegraphics[width=\textwidth]{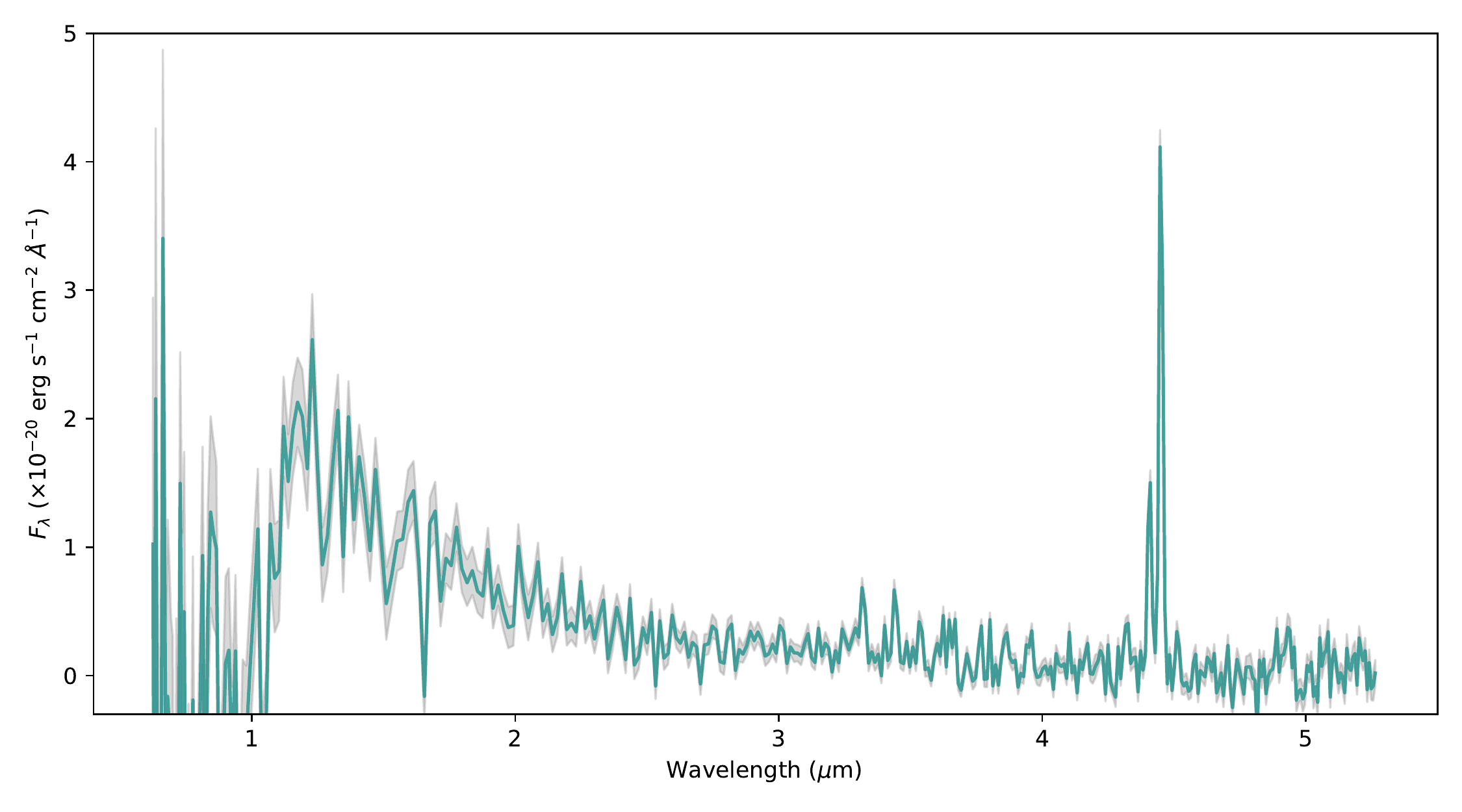} 
\caption{Abell-$z7885$} 
\end{subfigure}
 \hfill
\begin{subfigure}[t]{0.45\textwidth}
\centering
\includegraphics[width=\textwidth]{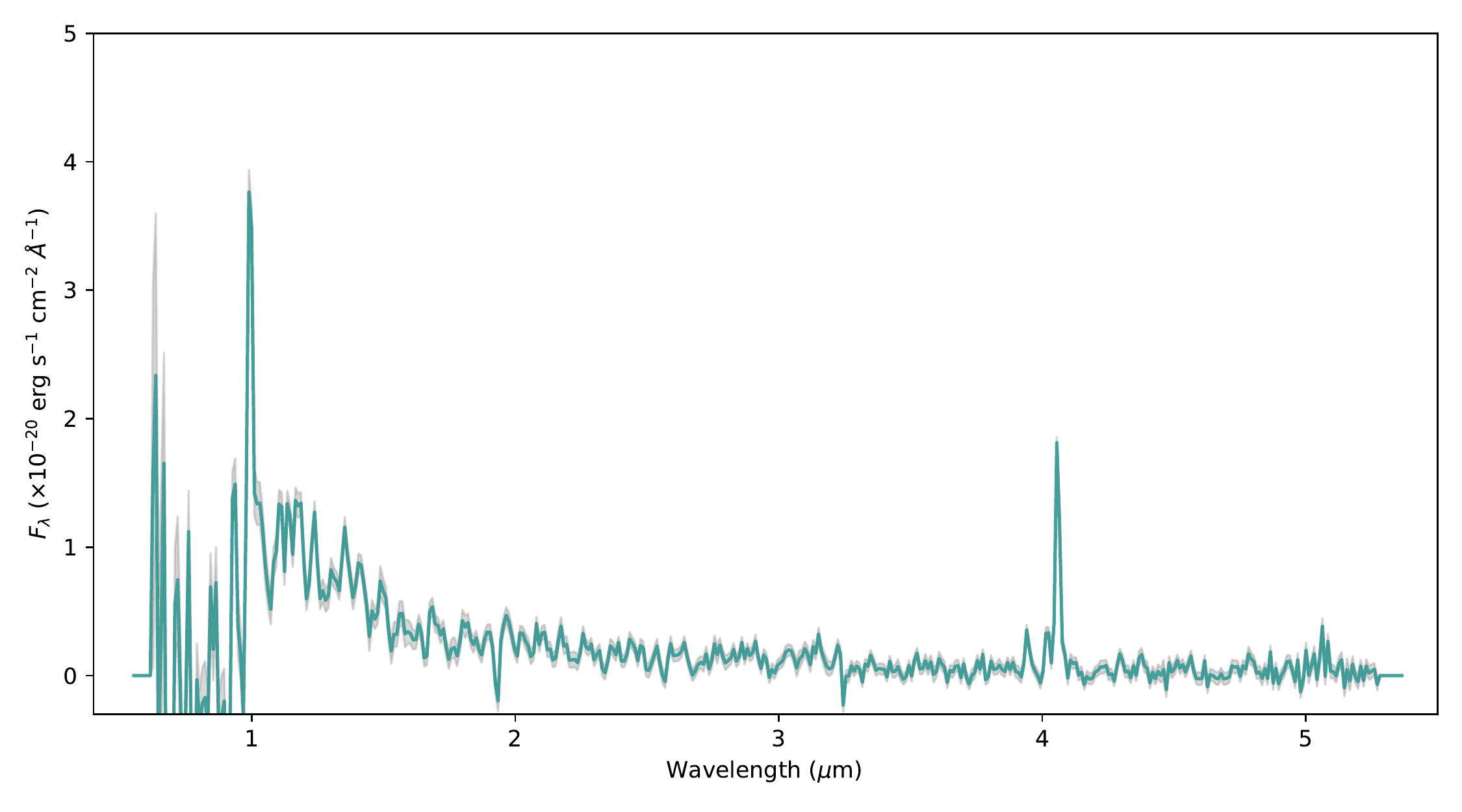} 
\caption{CEERS-$z7885$} 
\end{subfigure}
 \hfill
\begin{subfigure}[t]{0.45\textwidth}
\centering
\includegraphics[width=\textwidth]{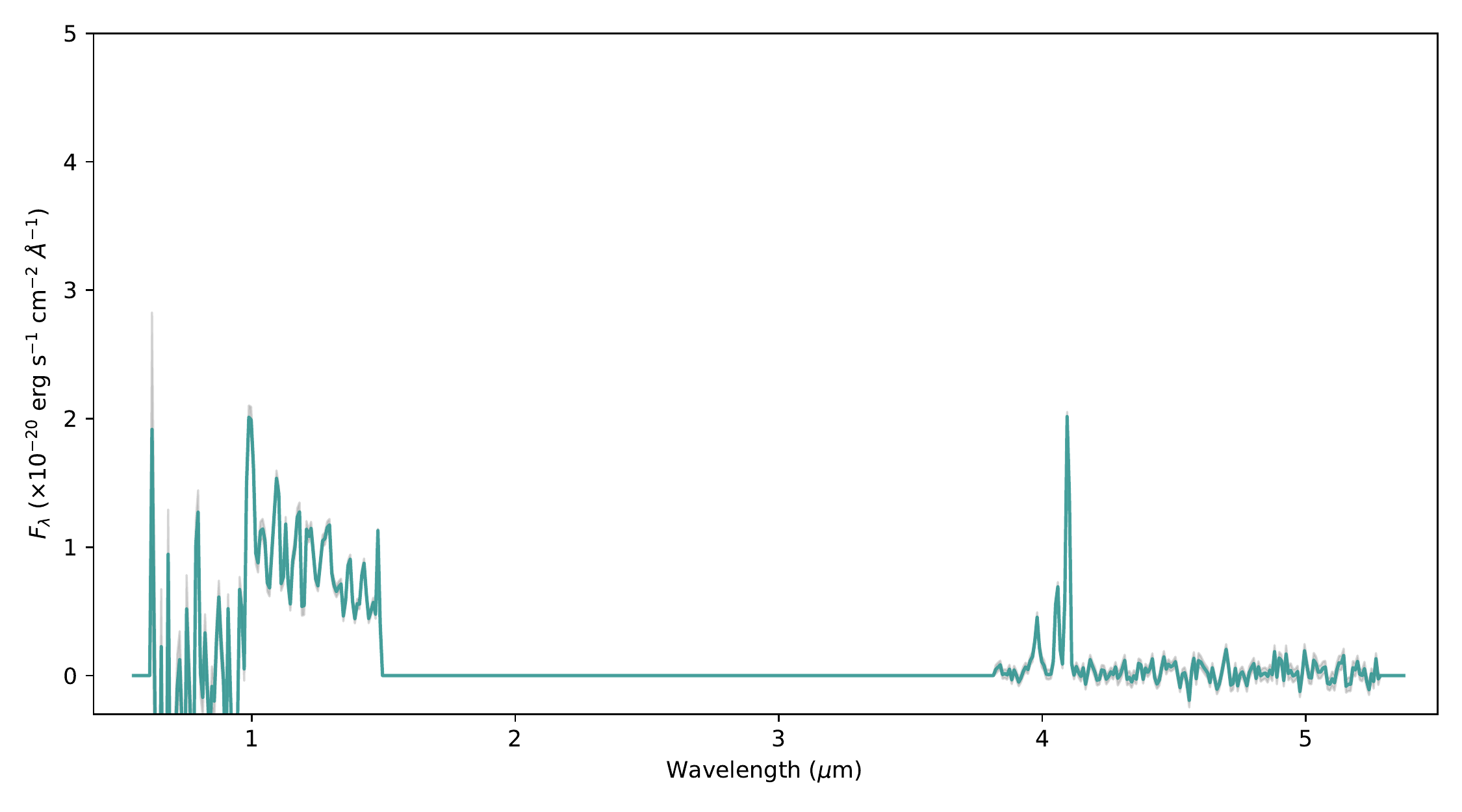} 
\caption{CEERS-$z7179$} 
\end{subfigure}
 \hfill
\begin{subfigure}[t]{0.45\textwidth}
\centering
\includegraphics[width=\textwidth]{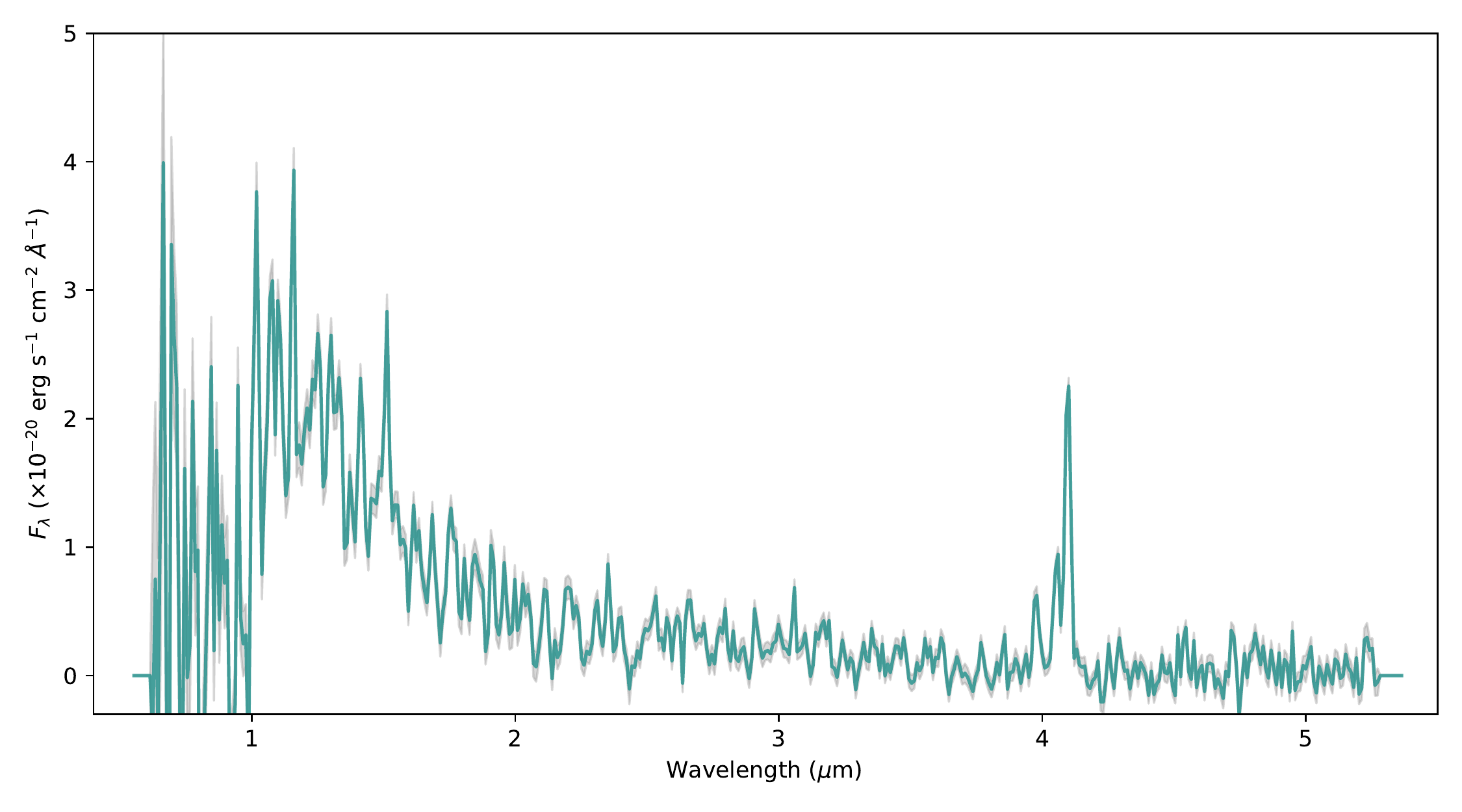} 
\caption{CEERS-$z7175$} 
\end{subfigure}
\caption{Photometrically-calibrated 1D NIRSpec/prism spectra of the full sample of galaxies at $z=7-10$.}
\label{efig:allspec}
\end{figure}

\begin{figure}
\centering
\begin{subfigure}[t]{0.45\textwidth}
\centering
\includegraphics[width=\textwidth]{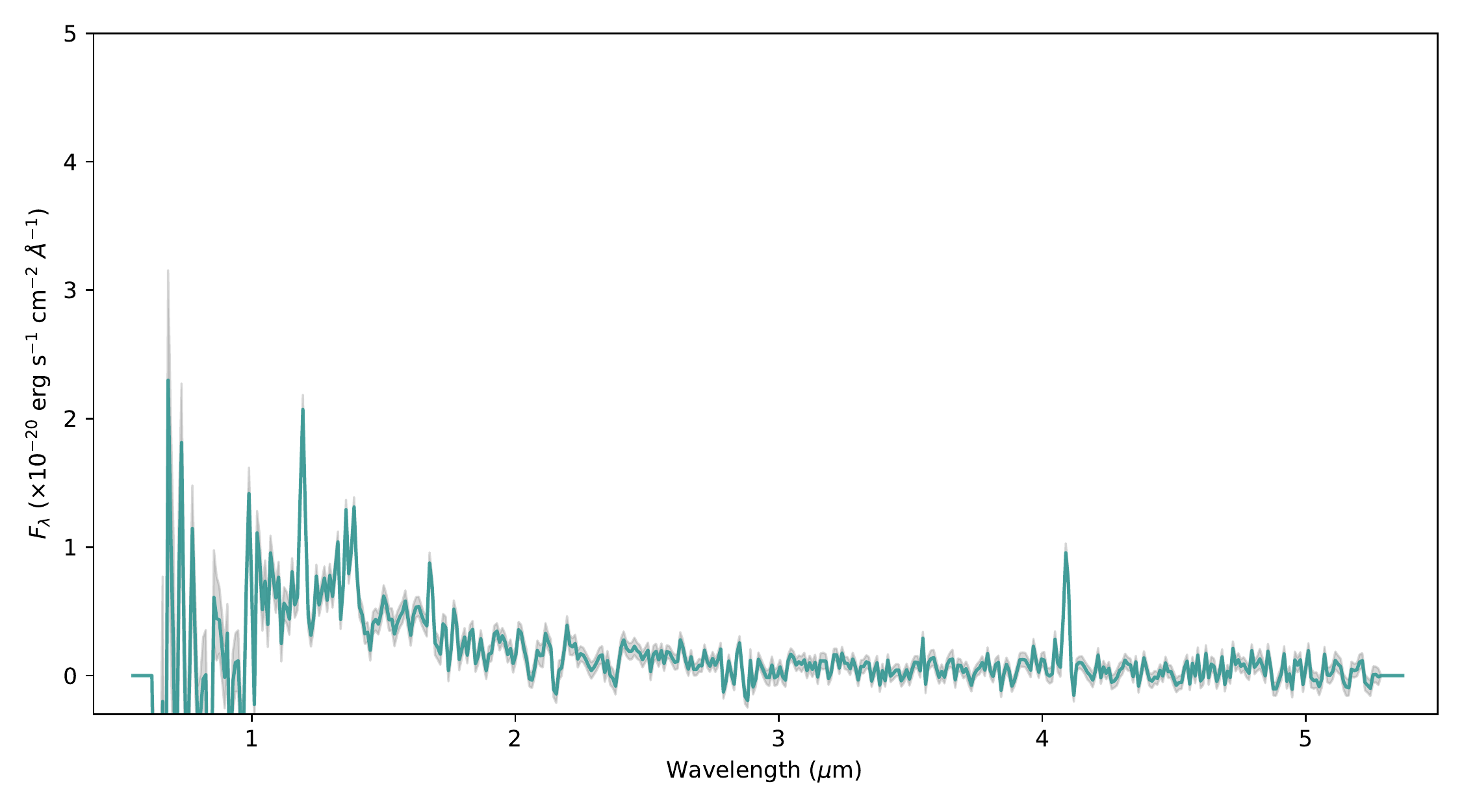} 
\caption{CEERS-$z7167$} 
\end{subfigure}
 \hfill
\begin{subfigure}[t]{0.45\textwidth}
\centering
\includegraphics[width=\textwidth]{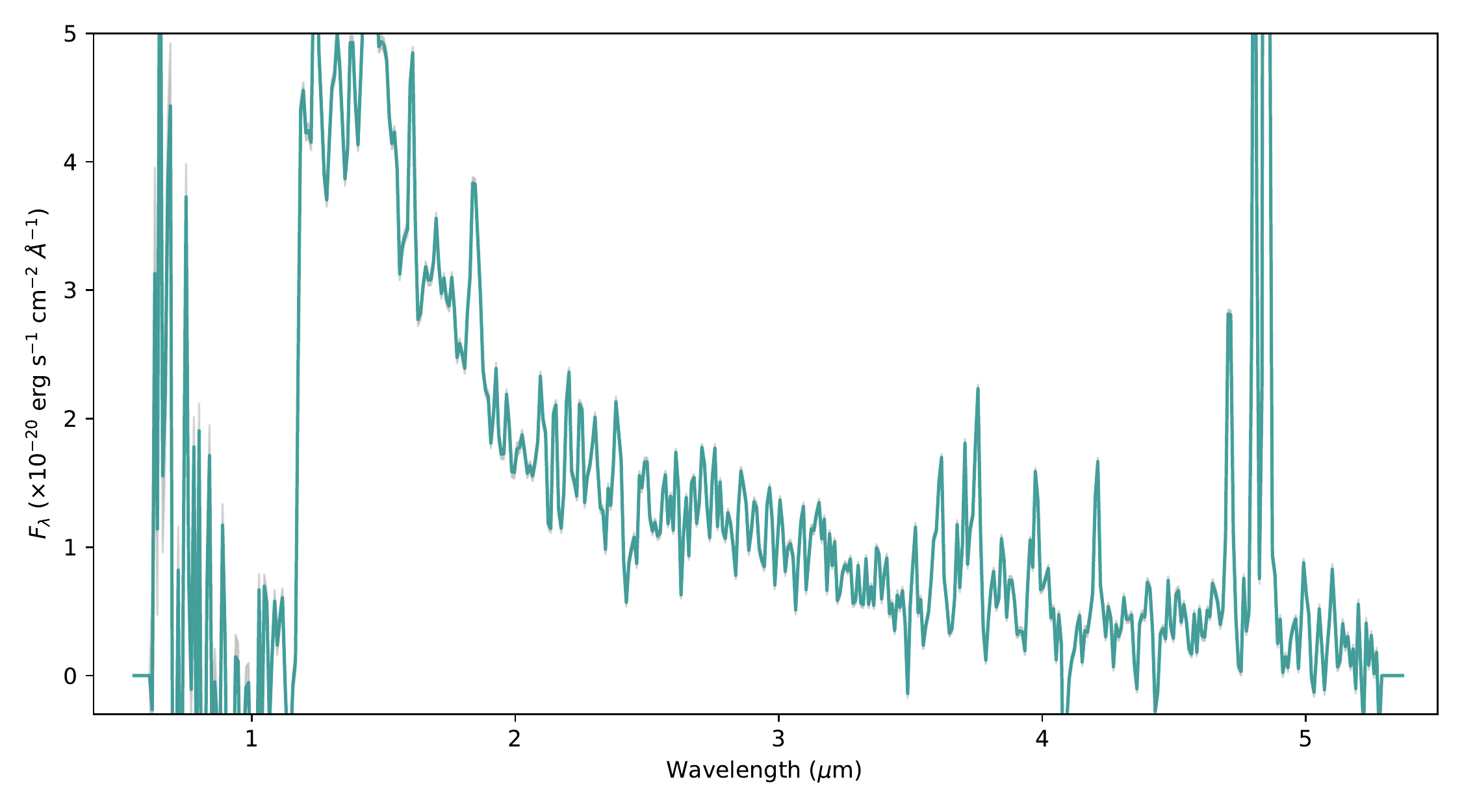} 
\caption{CEERS-$z8684$} 
\end{subfigure}
\hfill
\begin{subfigure}[t]{0.45\textwidth}
\centering
\includegraphics[width=\textwidth]{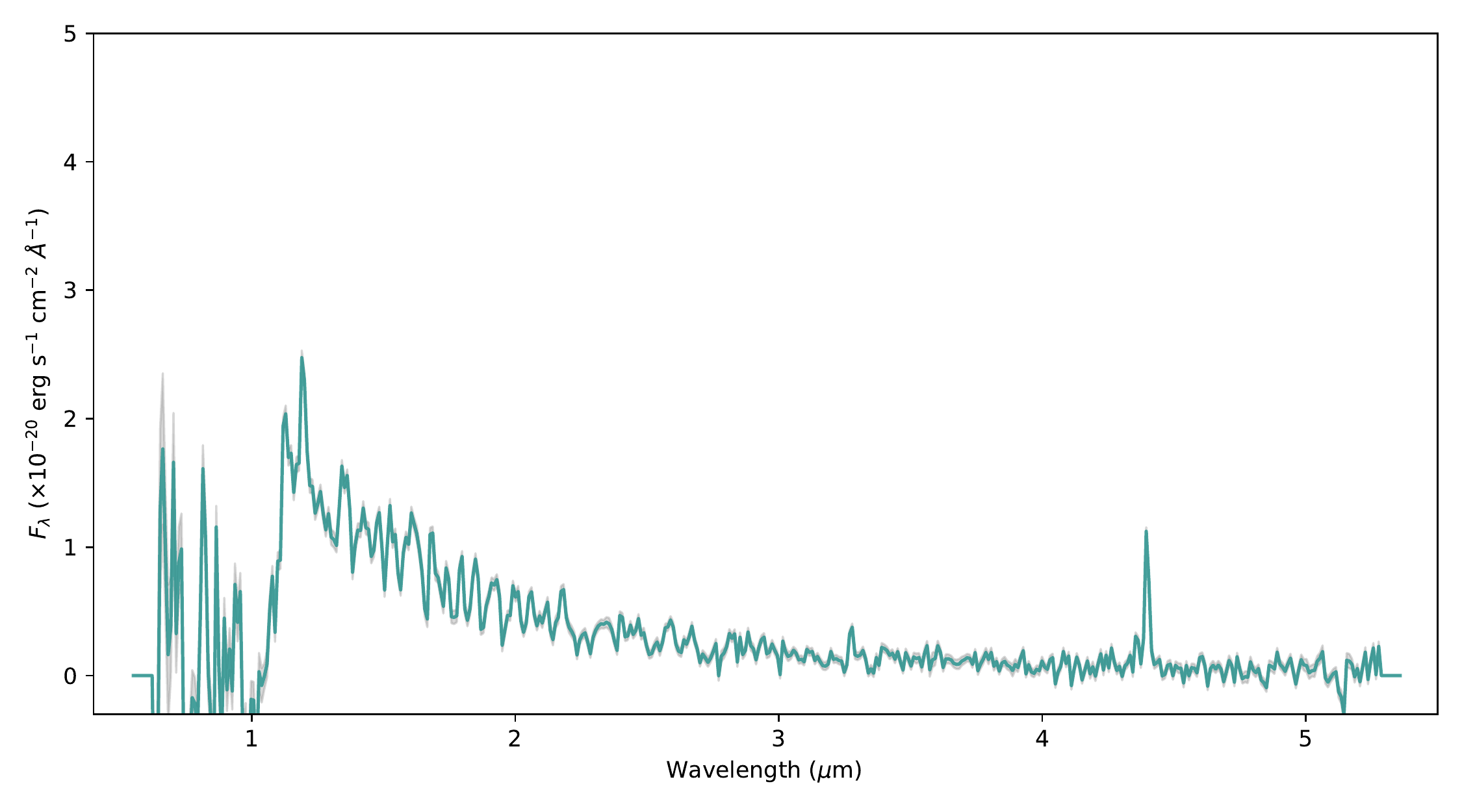} 
\caption{CEERS-$z7789$} 
\end{subfigure}
 \hfill
\begin{subfigure}[t]{0.45\textwidth}
\centering
\includegraphics[width=\textwidth]{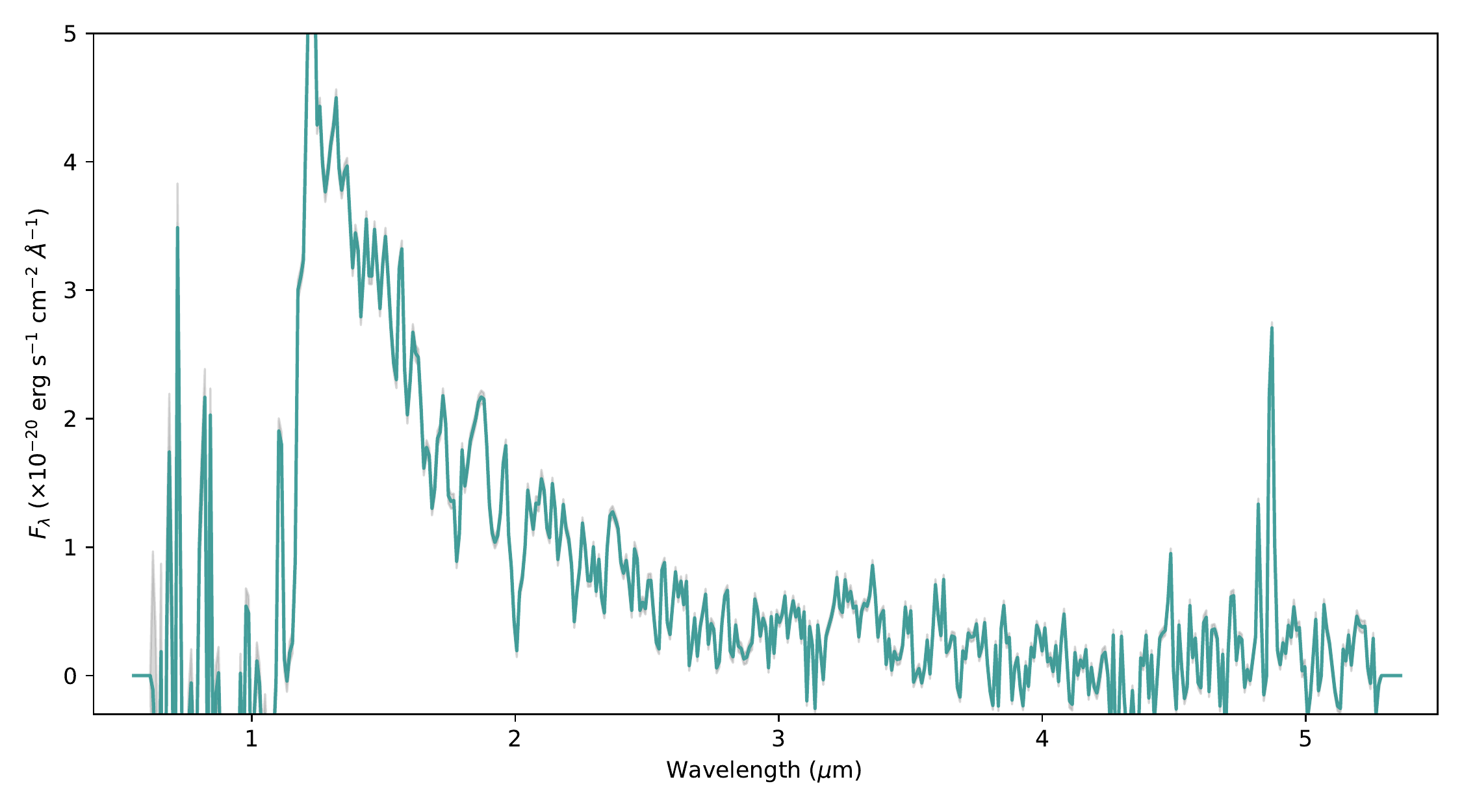} 
\caption{CEERS-$z8718$} 
\end{subfigure}
 \hfill
\begin{subfigure}[t]{0.45\textwidth}
\centering
\includegraphics[width=\textwidth]{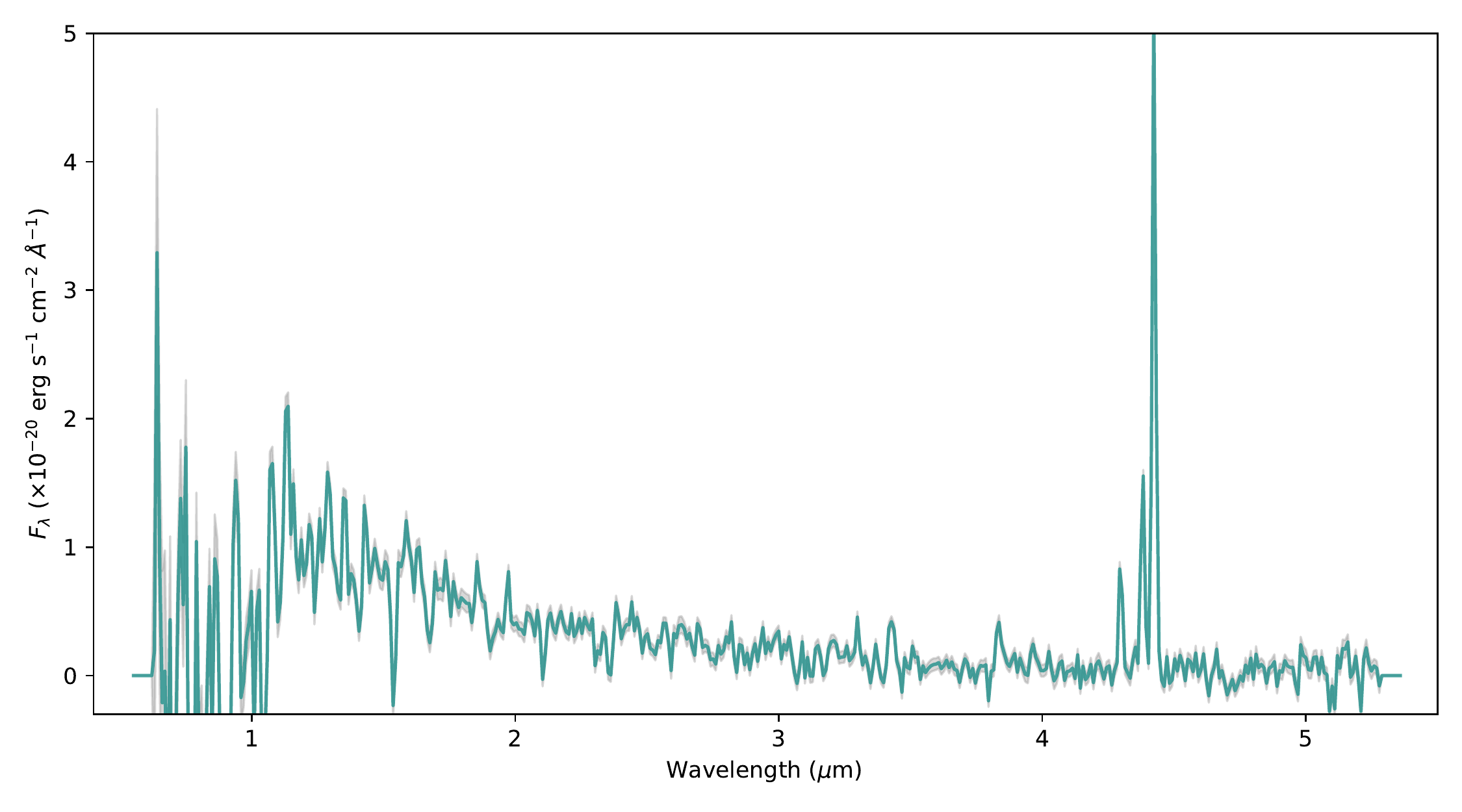} 
\caption{CEERS-$z7832$} 
\end{subfigure}
 \hfill
\begin{subfigure}[t]{0.45\textwidth}
\centering
\includegraphics[width=\textwidth]{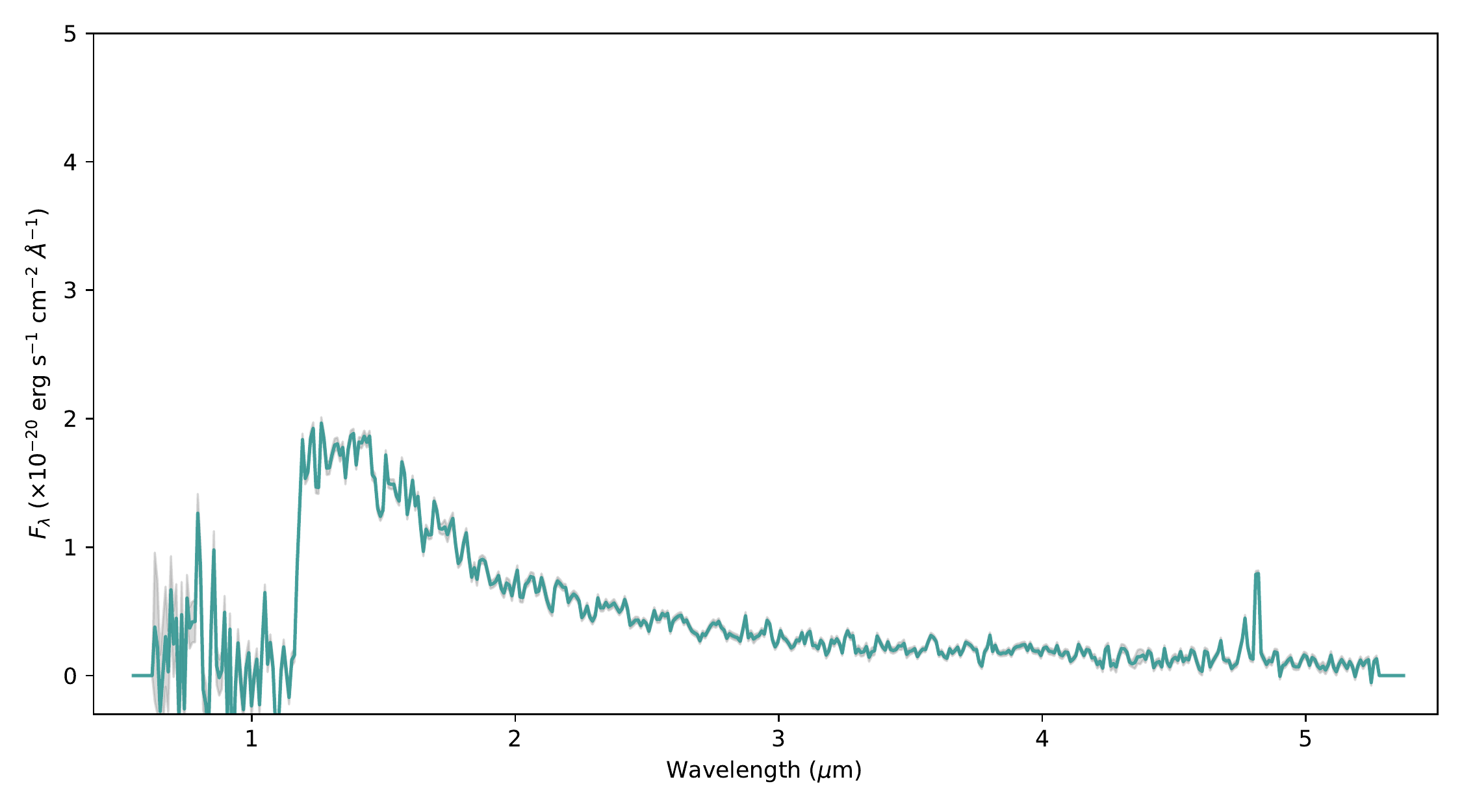} 
\caption{CEERS-$z8612$} 
\end{subfigure}
 \hfill
\begin{subfigure}[t]{0.45\textwidth}
\centering
\includegraphics[width=\textwidth]{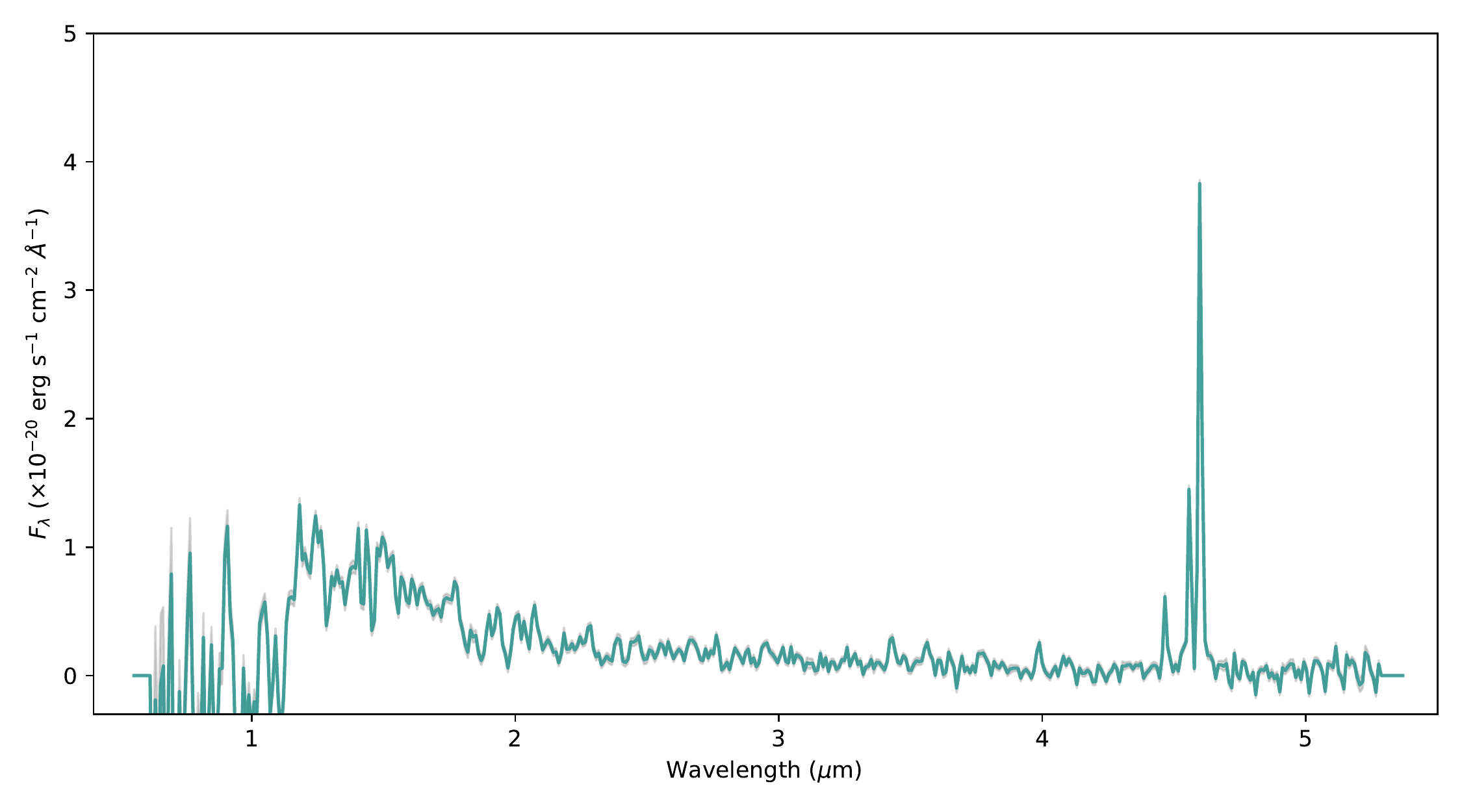} 
\caption{CEERS-$z8172$} 
\end{subfigure}
 \hfill
\begin{subfigure}[t]{0.45\textwidth}
\centering
\includegraphics[width=\textwidth]{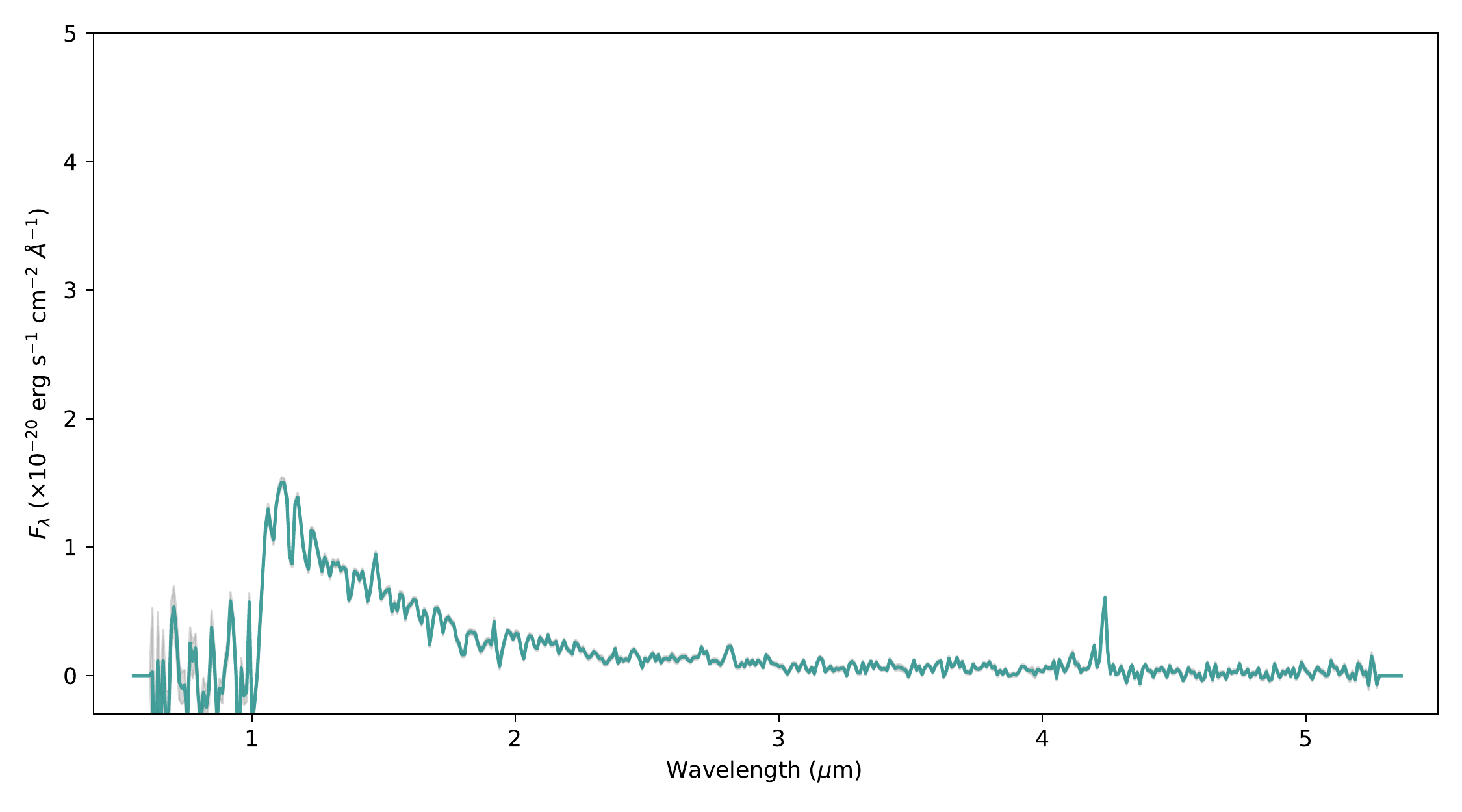} 
\caption{CEERS-$z7453$} 
\end{subfigure}
 \caption*{(Continued).}
\end{figure}

\begin{figure}
\centering
\includegraphics[width=0.95\textwidth]{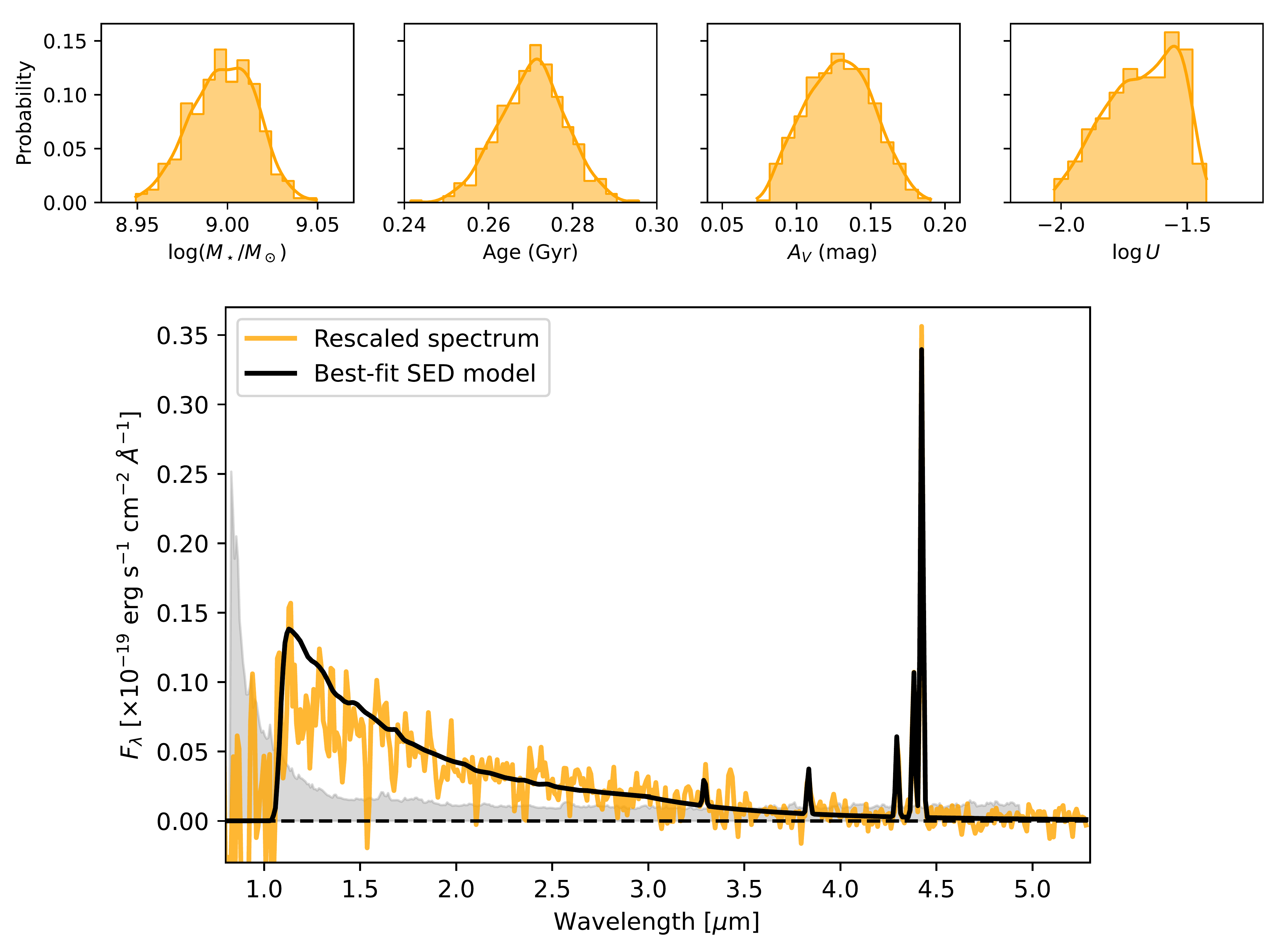}
\caption{The photometrically-calibrated 1D spectrum of the galaxy CEERS-$z7832$ (orange), imposed with the best-fit continuum and line emission from the {\sc Bagpipes} SED model (black). The grey region represents the uncertainty on the flux. The output posterior distributions from the best-fit SED on the stellar mass $M_\star$, mass-weighted age $\tau_{\rm mass}$, dust attenuation $A_V$, and ionization parameter $U$, are shown at the top. }
\label{efig:bpfit}
\end{figure}

\begin{table}[h]
\begin{center}
\begin{minipage}{300pt}
\caption{Line flux measurements.}
\begin{tabular}{@{}lccccc@{}}
\toprule
Galaxy ID  & $z_{\rm spec}$  & H$\beta$ & [\oiii]\,$\lambda 5007$ & [\oii]\,$\lambda 3727$ \\
\midrule
RXJ-z9500 & 9.5008 & $18.01\pm 1.61$ & $91.44\pm 1.90$ & $4.94\pm 1.87$  \\
RXJ-z8149 & 8.1496 & $5.76\pm 2.27$  & $19.62\pm 2.57$ & $<2.54$ \\
\vspace{0.1cm}
RXJ-z8152 & 8.1523 & $9.37\pm 1.90$  & $53.99\pm 2.23$ & $<5.21$ \\
Abell-z7878 & 7.8783 & $<8.01$  & $28.37\pm 4.58$ & $9.67\pm 4.55$ \\
\vspace{0.1cm}
Abell-z7885 & 7.8854 & $7.13\pm 2.27$  & $77.69\pm 2.67$ & $8.74\pm 2.66$ \\
CEERS-z7105 & 7.1051 & $5.27 \pm   1.23$  &  $33.72  \pm 1.44$  &  $<2.44$ \\
CEERS-z7179 & 7.1792 & $7.84 \pm   0.73$ &  $36.38 \pm  0.84$ & $--$ \\
CEERS-z7175 & 7.1751 & $14.46 \pm  2.64$ &    $60.91 \pm  2.86$ &   $7.54 \pm   2.66$ \\
CEERS-z7167 & 7.1683 & $3.01  \pm  1.38 $ &  $19.73 \pm  1.56$ &  $2.73 \pm   1.39$ \\
CEERS-z8684 & 8.6849 & $57.60 \pm  5.95$ &   $461.03 \pm 7.10$  &  $18.68 \pm  5.83$ \\
CEERS-z7789 & 7.7805 & $<2.61$ & $19.99\pm   1.23 $ &  $4.33\pm    1.05$ \\
CEERS-z8718 & 8.7182 & $12.27\pm   4.68$ &    $62.04  \pm 5.57$ &    $7.89\pm    4.58$ \\
CEERS-z7832 & 7.8328 & $14.30 \pm  1.60$ &   $87.42 \pm  1.86$  & $ 3.81\pm    1.62$\\
CEERS-z8612 & 8.6142 & $2.69  \pm  0.80  $  & $15.87 \pm   0.98$ &  $1.91  \pm  0.79$ \\
CEERS-z8172 & 8.1724 & $8.94  \pm  1.12$  & $64.09  \pm 1.35$ &  $3.50  \pm  1.12$ \\
CEERS-z7453 & 7.4536 & $2.82 \pm   0.66$  &  $11.85 \pm  0.76 $  & $1.10 \pm   0.66$ \\
\botrule
\end{tabular}
\footnotetext{{\bf Notes.} The derived line fluxes are in units of $10^{19}$\,erg\,s$^{-1}$\,cm$^{-2}$ and have not been corrected for the magnification factor $\mu$. Upper limits denote $2\sigma$ upper bounds on the line fluxes.}
\label{tab:lflux}
\end{minipage}
\end{center}
\end{table}

\begin{figure}
\centering
\includegraphics[width=0.8\textwidth]{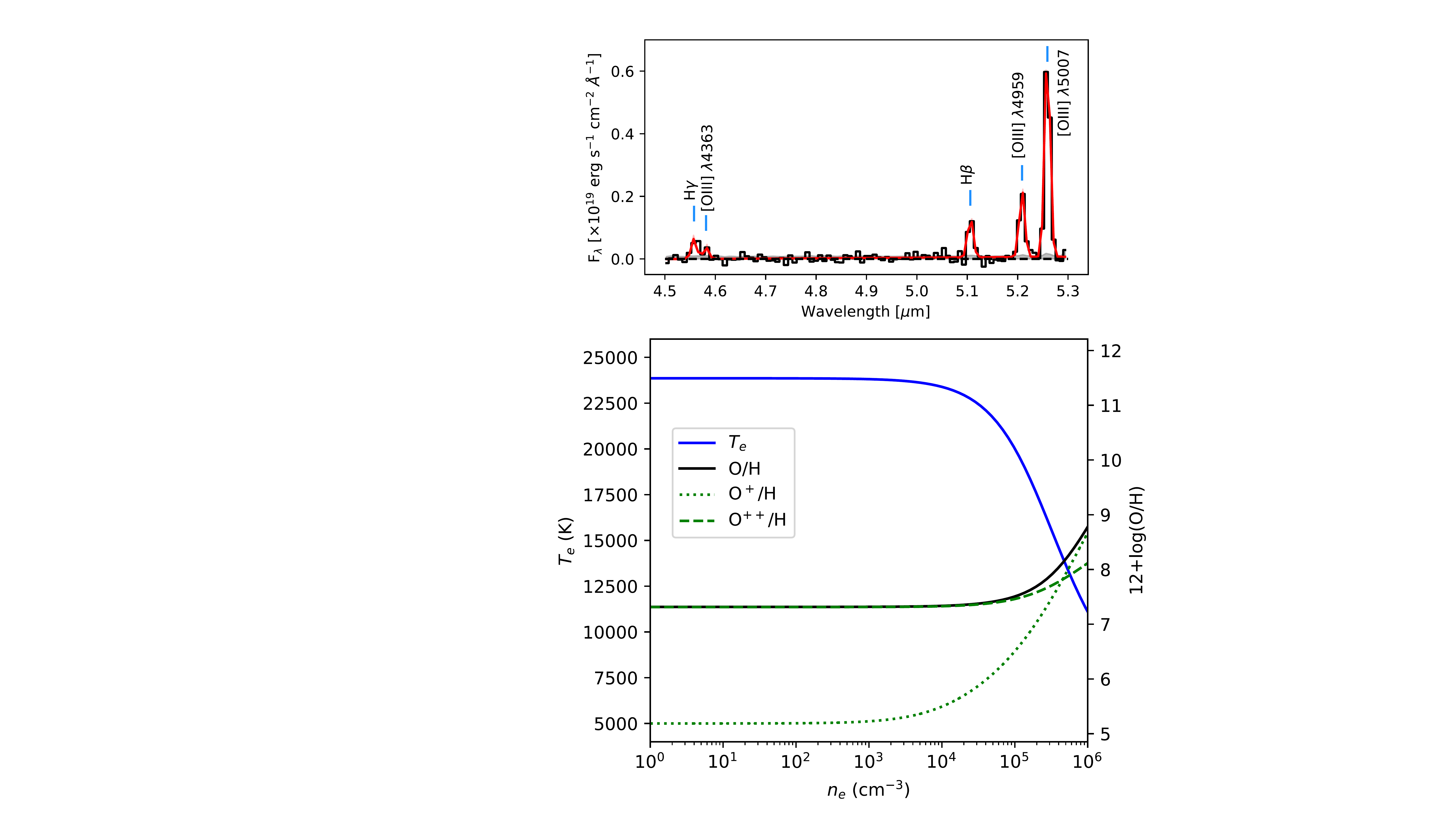}
\caption{(Top): Zoom-in on the photometrically-calibrated 1D spectrum of the galaxy  RXJ-$z9500$ (black), marking the most prominent emission line features. The typical strong nebular emission lines are detected, in addition to a marginal detection of the auroral [\oiii]\,$\lambda 4363$ transition. The best-fit model of the continuum and line profiles are shown in red. 
(Bottom): The derived electron temperature $T_e$ and oxygen abundance $12+\log$(O/H) for RXJ-$z9500$ as a function of electron density $n_e$, based on the direct $T_e$-method \cite{Izotov06}.}
\label{efig:auroralmet}
\end{figure}


\begin{figure}
\centering
\includegraphics[width=0.7\textwidth]{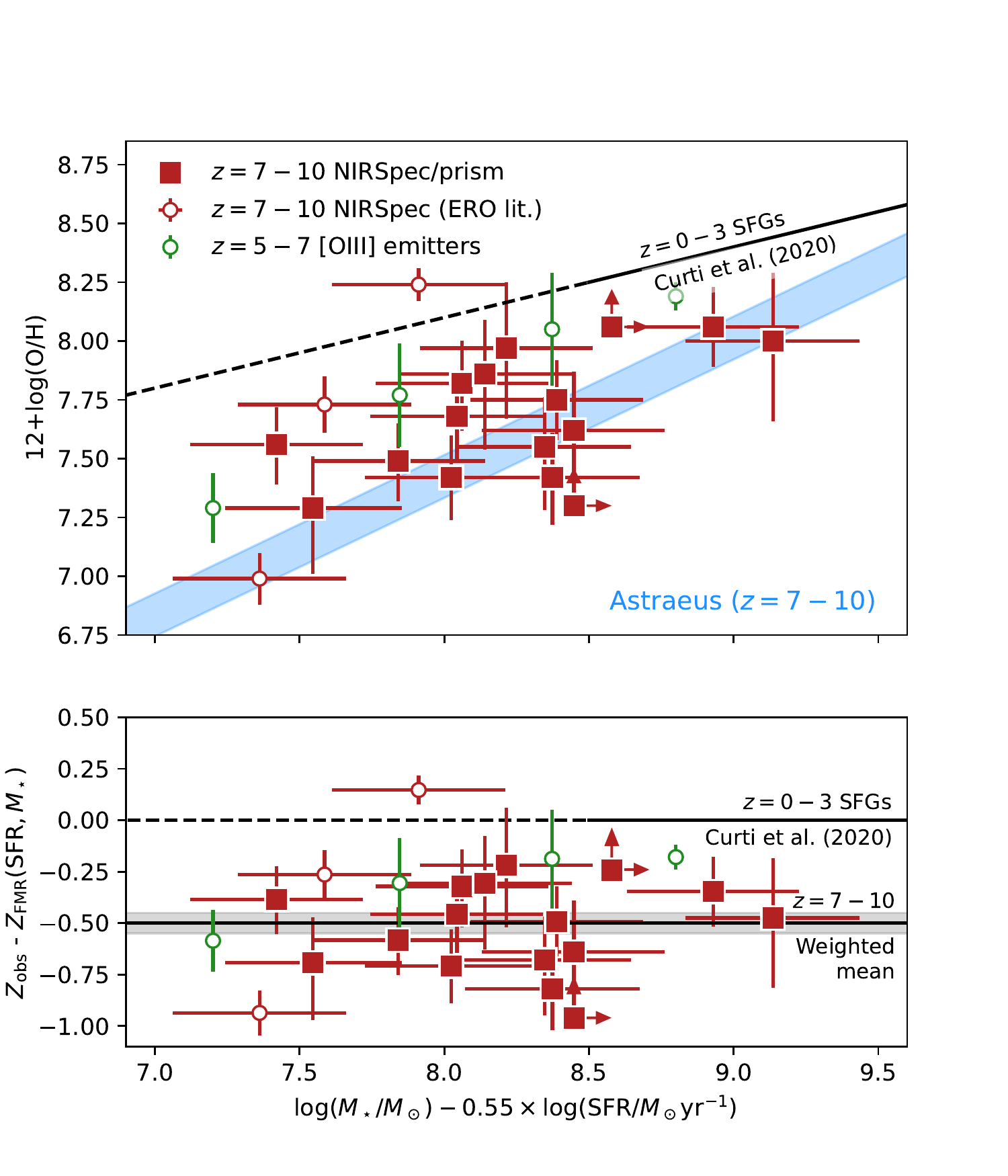}
\caption{The fundamental-metallicity relation (FMR) of galaxies, similar to Fig.~\ref{fig:fmr}. For comparison, we show the sample of [\oiii]-emitters at $z=5-7$ in different mass bins \cite{Matthee22}, and the three galaxies at $z>7$ with direct metallicity estimates from the $T_e$-method \cite{Curti23}.}
\label{efig:fmrcomp}
\end{figure}

\clearpage
\newpage

\setstretch{1.}

\newcommand{\mnras}{Monthly Noticies of the Royal Astronomical Society}
\newcommand{\aap}{Astronomy \& Astrophysics}
\newcommand{\apj}{Astrophysical Journal}
\newcommand{\apjl}{Astrophysical Journal Letters}
\newcommand{\apjs}{Astrophysical Journal Supplement Series}
\newcommand{\araa}{Annual Review of Astronomy and Astrophysics}
\newcommand{\aapr}{Astronomy and Astrophysics Review}
\newcommand{\rmxaa}{Revista Mexicana de Astronomía y Astrofísica}
\newcommand{\pasp}{Publications of the Astronomical Society of the Pacific}
\newcommand{\physrep}{Physics Reports}

\bibliographystyle{apj}
\bibliography{ref.bib}


\end{document}